\documentclass[a4paper,11pt]{article}
\usepackage{pos}
\usepackage{mathtools}
\usepackage[makeroom]{cancel}
\usepackage{bbold}

\title{Unification of Gravity with Internal Interactions: the SO(2,16) Scheme Revisited}
\author*[a]{Stelios Stefas}
\author[a,b,c,d]{George Zoupanos}

\affiliation[a]{Physics Department, National Technical University of Athens, Zografou Campus,\\
157 80, Zografou, Greece}

\affiliation[b]{Max-Planck Institut f\"ur Physik, Boltzmannstr. 8,\\
85 748 Garching/Munich, Germany}

\affiliation[c]{Universit\"at Hamburg, Luruper Chaussee 149,\\
22 761 Hamburg, Germany}

\affiliation[d]{Deutsches Elektronen-Synchrotron DESY, Notkestra{\ss}e 85,\\
22 607, Hamburg, Germany}

\emailAdd{dstefas@mail.ntua.gr}
\emailAdd{george.zoupanos@cern.ch}

\abstract{The starting point of the present work is the observation that the tangent group of a curved manifold need not have the same dimension as the manifold itself. We use this fact to construct Conformal Gravity and its noncommutative (Fuzzy) counterpart as gauge theories, and then to unify both with internal gauge interactions. Specifically, we present a scheme in which conformal gravity, based on gauging the $SO(2,4)$ group, and fuzzy gravity, based on gauging the $SO(1,5)\times U(1)$, are unified with $SO(10)$ internal interactions through the parent group $SO(2,16)$, and we work through the resulting spontaneous symmetry breakings and fermion content in detail. We also discuss the low-energy phenomenology of the scheme, including cosmic-string gravitational-wave signals, and close with a brief outlook: a candidate minimal unification group, a dark-matter candidate suggested by the construction, and the ghost problem common to higher-derivative gravity theories of this type.}

\FullConference{Proceedings of the Corfu Summer Institute 2025 "School and Workshops on Elementary Particle Physics and Gravity" (CORFU2025)\\
27 April - 28 September, 2025\\
Corfu, Greece\\}

\begin{document}
\maketitle

\section{Introduction}
\label{sec1}

The unification of all fundamental interactions has been pursued for over a century, since Kaluza and Klein \cite{Kaluza:1921, Klein:1926} first showed that adding a fifth dimension to spacetime allows gravity and electromagnetism to be described within a single geometric framework. This was later generalized to several extra dimensions, and it was found \cite{Kerner:1968, CHO1987358, Cho:1975sf} that non-Abelian gauge symmetries of the type needed for the Standard Model (SM) arise naturally whenever more than one extra dimension is present. In particular, if the full spacetime is taken to be $M_D = M_4 \times B$, with $B$ a compact Riemannian manifold with non-Abelian isometry group $S$, dimensional reduction of pure gravity in $D$ dimensions produces, in four dimensions, ordinary gravity coupled to a Yang-Mills theory with gauge group $S$, along with a set of scalar fields. Gauge symmetry thus arises purely from geometry. Two problems remain, however: there is in general no stable ground state of the required product form, and chiral fermions cannot be obtained in four dimensions through this type of reduction \cite{Witten:1983}.

A way to circumvent the second problem is to introduce Yang-Mills fields in the higher-dimensional theory from the outset, at the expense of the purely geometric character of the construction. If both gauge fields and fermions are present in a higher-dimensional Grand Unified Theory (GUT) \cite{Georgi1999, FRITZSCH1975193}, chirality in four dimensions requires the total spacetime dimension to be of the form $4k+2$ (see e.g.~\cite{CHAPLINE1982461}). In the present work we adopt a different point of view, in which every fundamental interaction, including gravity, is treated as arising from the gauging of a symmetry, rather than from geometry, compactification, or dimensional reduction. Historically, however, the dominant framework for higher-dimensional unification has been Superstring theory \cite{Green2012-ul, polchinski_1998, Lust:1989tj}.

The heterotic string \cite{GROSS1985253}, formulated consistently in ten dimensions, contains GUT-scale gauge groups such as $E_8\times E_8$, from which the SM can in principle be obtained. However, a separate, purely field-theoretic approach to higher-dimensional model building had already been developed before Superstring theory became dominant \cite{forgacs, MANTON1981502, Kubyshin:1989vd, KAPETANAKIS19924, LUST1985309}. Forgacs and Manton (F-M) introduced the Coset Space Dimensional Reduction (CSDR) scheme \cite{forgacs, MANTON1981502, Kubyshin:1989vd, KAPETANAKIS19924}, which naturally leads to chiral fermions, while Scherk and Schwarz (S-S) proposed reduction on a group manifold \cite{SCHERK197961}; the latter does not yield chirality but influenced later work in string phenomenology. Model building along these CSDR lines, with a view to realistic phenomenology, is pursued further in \cite{Manousselis_2004, Chatzistavrakidis:2009mh, Irges:2011de, Manolakos:2020cco, Patellis:2024dfl}.

A different route to unification stays entirely within four dimensions, relying on the fact that both the SM and gravity can be formulated as gauge theories \cite{utiyama, kibble1961, Sciama, Umezawa, Matsumoto, macdowell, Ivanov:1980tw, Ivanov:1981wn, stellewest, Kibble:1985sn}. This approach predates Superstring theory and received renewed attention through supergravity \cite{freedman_vanproeyen_2012, Ortín_2015}, which is itself a gauge theory, and, more recently, through its extension to noncommutative gravity \cite{castellani, Chatzistavrakidis_2018, Manolakos_paper1, manolakosphd, Manolakos_paper2, Manolakos:2022universe, Manolakos:2023hif, roumelioti2407}.

This approach goes back to Weyl \cite{weyl, weyl1929}, who related electromagnetism to local phase transformations of the electron field and, in doing so, introduced the vierbein required by any gauge formulation of gravity. Utiyama \cite{utiyama} showed that gravity can be understood as the gauge theory of the Lorentz group $SO(1,3)$, although the introduction of the vierbein in his treatment was somewhat ad hoc; Kibble \cite{kibble1961} and Sciama \cite{Sciama} removed this ad hoc element by gauging the full Poincar\'e group instead. Stelle and West \cite{stellewest, Kibble:1985sn} showed that gauging the de Sitter ($SO(1,4)$) or anti-de Sitter ($SO(2,3)$) group, followed by spontaneous breaking to the Lorentz group, provides an even more economical starting point than the Poincar\'e group (see also \cite{Addazi:2024xkg}). Gauging the larger, fifteen-generator conformal group $SO(2,4)$ leads to Weyl Gravity (WG) \cite{KAKU1977304, Roumelioti:2024lvn} and to Fuzzy Gravity (FG) \cite{Chatzistavrakidis_2018, Manolakos_paper1, manolakosphd, Manolakos_paper2, Manolakos:2022universe, Manolakos:2023hif, roumelioti2407}, both of which admit $\mathcal{N}=1$ supergravity extensions \cite{freedman_vanproeyen_2012}.

A more ambitious step is to place gravity and the internal gauge symmetries of particle physics inside a single group \cite{Percacci:1984ai, Percacci_1991}, an idea revisited in a number of recent works \cite{Nesti_2008, Nesti_2010, Chamseddine2010, Chamseddine2016, Krasnov:2017epi, Konitopoulos:2023wst, Manolakos:2023hif, noncomtomos, Roumelioti:2024lvn, Patellis:2024znm, Roumelioti:2025cxi,Patellis:2025qbl}. This is possible because the tangent group of a curved spacetime need not match the dimension of the manifold: a tangent group larger than four can therefore be gauged over ordinary four-dimensional spacetime, leaving room for internal symmetries alongside gravity. Several CSDR techniques developed for genuinely higher-dimensional theories \cite{CHAPLINE1982461, forgacs, MANTON1981502, Kubyshin:1989vd, KAPETANAKIS19924, LUST1985309, SCHERK197961, Manousselis_2004, Chatzistavrakidis:2009mh, Irges:2011de, Manolakos:2020cco, Patellis:2024dfl} carry over in the current scheme too, and so do the associated difficulties: a realistic chiral spectrum again requires imposing Weyl and Majorana conditions together \cite{CHAPLINE1982461, KAPETANAKIS19924}.

Along these lines, unification of the conformal gauge group with internal interactions was achieved in \cite{Manolakos:2023hif, Konitopoulos:2023wst, Patellis:2024znm, Roumelioti:2024lvn, Roumelioti:2025cxi,Patellis:2025qbl}, and subsequently extended to the noncommutative (fuzzy) case in \cite{roumelioti2407}. The $SO(2,16)$ scheme discussed here was first presented in our earlier proceedings contribution \cite{Roumelioti:2025nku}.

The rest of this contribution is organized as follows. We first review the gauge-theoretic construction of gravity in general, then construct Conformal Gravity and Noncommutative (Fuzzy) Gravity within this language, and finally enlarge the tangent group up to $SO(2,16)$ so as to incorporate internal interactions. We then discuss the resulting low-energy phenomenology, and close with an outlook covering a candidate minimal unification group compatible with a simultaneous Weyl-Majorana condition, a dark-matter candidate, and the ghost problem common to higher-derivative gravity theories of this type.

\section{Gauge gravity of the Conformal Group, $SO(2,4)$}

Einstein Gravity (EG) is usually obtained by gauging the Poincar\'e group. A more economical and instructive starting point, however, is provided by the ten-generator de Sitter group $SO(1,4)$ or its anti-de Sitter counterpart $SO(2,3)$, either of which reduces to the Lorentz group $SO(1,3)$ through spontaneous breaking driven by non-dynamical, auxiliary scalar fields \cite{stellewest, Kibble:1985sn, Roumelioti:2024lvn, manolakosphd}. The Poincar\'e, dS, and AdS groups are all subgroups of the fifteen-generator conformal group $SO(2,4)$, which leaves the null interval $ds^2=\eta_{\mu\nu}dx^\mu dx^\nu=0$ invariant. Gauging $SO(2,4)$ directly gives Conformal Gravity (CG) \cite{Kaku:1978nz}; in that original construction, the reduction from CG to either EG or to Weyl's scale-invariant gravity was implemented through constraints imposed by hand on the gauge fields. Below we instead follow \cite{Roumelioti:2024lvn}, where the same reduction is achieved dynamically, through spontaneous breaking of the conformal gauge symmetry induced by a scalar field acquiring a vacuum expectation value (vev) via a Lagrange multiplier term in the action.

\subsection{Spontaneous symmetry breaking by introducing a scalar in the adjoint representation}
In four dimensions the fifteen $SO(2,4)$ generators organize into six Lorentz generators $M_{ab}$, four translation generators $P_a$, four special-conformal (conformal-boost) generators $K_a$, and one dilatation generator $D$.

The associated gauge connection $A_\mu$ takes the form

\begin{equation}
A_\mu= \frac{1}{2}\omega_\mu{}^{a b} M_{a b}+e_\mu{}^a P_a+b_\mu{}^a K_a+\tilde{a}_\mu D,
\end{equation}
where the translation-sector field plays the role of vierbein and the Lorentz-sector field that of the spin connection; its field strength is
\begin{equation}\label{fst}
F_{\mu \nu}=\frac{1}{2}R_{\mu \nu}{}^{a b} M_{a b}+\tilde{R}_{\mu \nu}{}^a P_a+R_{\mu \nu}{}^a K_a+R_{\mu \nu} D,
\end{equation}
with components
\begin{equation}\label{curves}
\begin{aligned}
R_{\mu \nu}{}^{a b} & =\partial_\mu \omega_\nu{}^{a b}-\partial_\nu \omega_\mu{}^{a b}-\omega_\mu{}^{a c} \omega_{\nu c}{}^b+\omega_\nu{}^{a c} \omega_{\mu c}{}^b-8 e_{[\mu}{}^{[a} b_{\nu]}{}^{b]} \\
& =R_{\mu \nu}^{(0) a b}-8 e_{[\mu}{}^{[a} b_{\nu]}{}^{b]}, \\
\tilde{R}_{\mu \nu}{}^a & =\partial_\mu e_\nu{}^a-\partial_\nu e_\mu{}^a+\omega_\mu{}^{a b} e_{\nu b}-\omega_\nu{}^{a b} e_{\mu b}-2 \tilde{a}_{[\mu} e_{\nu]}{}^a \\
& =T_{\mu \nu}^{(0) a}(e)-2 \tilde{a}_{[\mu} e_{\nu]}{}^a, \\
R_{\mu \nu}{}^a & =\partial_\mu b_\nu{}^a-\partial_\nu b_\mu{}^a+\omega_\mu{}^{a b} b_{\nu b}-\omega_\nu{}^{a b} b_{\mu b}+2 \tilde{a}_{[\mu} b_{\nu]}{}^a\\
&=T_{\mu \nu}^{(0) a}(b)+2 \tilde{a}_{[\mu} b_{\nu]}{}^a,\\
R_{\mu \nu} & =\partial_\mu \tilde{a}_\nu-\partial_\nu \tilde{a}_\mu+4 e_{[\mu}{}^a b_{\nu] a},
\end{aligned}
\end{equation}
where $T_{\mu \nu}^{(0) a}(e)$ and $R_{\mu \nu}^{(0) a b}$ are the standard vierbein-formalism torsion and curvature of General Relativity (GR), and $T_{\mu \nu}^{(0) a}(b)$ is the torsion built from the second, $b_\mu{}^a$, vielbein-like field.

Building a parity-preserving action quadratic in \eqref{fst} calls for an auxiliary scalar $\phi$ living in the adjoint ($\mathbf{15}$-dimensional) representation of $SU(4)\sim SO(6)\sim SO(2,4)$, together with a mass scale $m$:
\begin{equation}
    S_{SO(2,4)}=a_{CG}\int d^4x [\operatorname{tr} \epsilon^{\mu \nu \rho \sigma} m\phi F_{\mu \nu}F_{\rho \sigma}+(\phi^2-m^{-2} \mathbb{1}_4)],
\end{equation}
the trace running over the gauge indices. As an adjoint field, $\phi$ expands as
\begin{equation}
\phi=\phi^{a b} M_{a b}+\tilde{\phi}^a P_a+\phi^a K_a+\tilde{\phi} D,
\end{equation}

Following \cite{Li:1973mq}, we choose the gauge in which $\phi$ becomes diagonal, $\operatorname{diag}(1,1,-1,-1)$, that is, pointing entirely along the dilatation generator:
\begin{equation}
    \phi=\phi^0=\tilde{\phi}D \xrightarrow{\phi^2=m^{-2}\mathbb{1}_4}\phi=-2m^{-1} D.
\end{equation}
With this choice the action reduces to
\begin{equation}
    S=-2a_{CG}\int d^4x \operatorname{tr} \epsilon^{\mu \nu \rho \sigma} F_{\mu \nu}F_{\rho \sigma}D.
\end{equation}
This gauge fixing is itself the symmetry-breaking step; rescaling $e$, $b$, and $\tilde{a}$ by $m$; expanding $F_{\mu\nu}$ and applying the (anti)commutators of the generators then gives
\begin{equation}
\begin{gathered}
    S=-2a_{CG}\int d^4x \operatorname{tr} \epsilon^{\mu \nu \rho \sigma}\Big[\frac{1}{4}R_{\mu \nu}{}^{ab}R_{\rho \sigma}{}^{cd}M_{ab}M_{cd}D+\\
    +i\epsilon_{abcd}(R_{\mu \nu}{}^{ab}R_{\rho \sigma}{}^{c} K^d D - R_{\mu \nu}{}^{ab}\tilde{R}_{\rho \sigma}{}^{c}P^{d}D)+\\+(\frac{1}{2}\tilde{R}_{\mu \nu}{}^{a}R_{\rho \sigma} + 2\tilde{R}_{\mu \nu}{}^{a}R_{\rho \sigma}{}^{b})M_{ab}+\\
    +(\frac{1}{4}R_{\mu \nu}R_{\rho \sigma}- 2\tilde{R}_{\mu \nu}{}^{a}R_{\rho \sigma a})D
    \Big].
\end{gathered}
\end{equation}
Using the trace identities
\begin{equation}
\begin{gathered}
   \operatorname{tr}[K^{d}D]=\operatorname{tr}[P^{d}D]=\operatorname{tr}[M_{ab}]= \operatorname{tr}[D]=0, \\
        \text{and}\quad \operatorname{tr}[M_{ab}M_{cd}D]=-\frac{1}{2}\epsilon_{abcd},
\end{gathered}
\end{equation}
everything but the Lorentz-invariant term drops out, leaving
\begin{equation}
\label{BrokenActionConformal}
     S_{\mathrm{SO}(1,3)}=\frac{a_{CG}}{4}\int d^4x \epsilon^{\mu \nu \rho \sigma}\epsilon_{abcd}R_{\mu \nu}{}^{ab}R_{\rho \sigma}{}^{cd},
\end{equation}
which no longer contains $\tilde{a}_\mu$, so we are free to set $\tilde{a}_\mu=0$; the $P$- and $K$-sector field strengths then reduce to
\begin{equation}
\begin{aligned}
  \tilde{R}_{\mu \nu}{}^a & =mT_{\mu \nu}^{(0) a}(e)-2 m^2\tilde{a}_{[\mu} e_{\nu]}{}^a \longrightarrow mT_{\mu \nu}^{(0) a}(e), \\
R_{\mu \nu}{}^a &=mT_{\mu \nu}^{(0) a}(b)+2m^2 \tilde{a}_{[\mu} b_{\nu]}{}^a \longrightarrow mT_{\mu \nu}^{(0) a}(b).
\end{aligned}
\end{equation}
As neither of these enters the action, both can consistently be dropped, $\tilde{R}_{\mu \nu}{}^a = R_{\mu \nu}{}^a = 0$, giving a torsion-free theory, and the same holds for $R_{\mu \nu}=0$, which via \eqref{curves} imposes the constraint
\begin{equation}\label{ef}
    e_\mu{}^a b_{\nu a}-e_{\nu}{}^{a}b_{\mu a}=0.
\end{equation}
Two particular solutions of this constraint are of interest here.

\subsubsection*{When $b_\mu{}^a = ae_\mu{}^a$ - Einstein-Hilbert (E-H) action in the presence of a cosmological constant}

Introduced originally in \cite{Chamseddine:2002fd}, this identification turns \eqref{BrokenActionConformal} into
\begin{equation}
\begin{aligned}
        S_{\mathrm{SO}(1,3)} =\frac{a_{CG}}{4}\int d^4 x \epsilon^{\mu \nu \rho \sigma} \epsilon_{a b c d}&\left[R_{\mu \nu}^{(0) a b}-4m^2a\left(e_\mu{}^a e_\nu{}^b-e_\mu{}^b e_\nu{}^a\right)\right]\\
       &\left[R_{\rho \sigma}^{(0) c d}-4m^2a\left(e_\rho{}^c e_\sigma{}^d-e_\rho{}^d e_\sigma{}^c\right)\right]
       \end{aligned}
\end{equation}
i.e.
\begin{equation}\label{so24finalaction}
\begin{aligned}
    S_{\mathrm{SO}(1,3)}=\frac{a_{CG}}{4}\int d^4 x \epsilon^{\mu \nu \rho \sigma} \epsilon_{a b c d}[R_{\mu \nu}^{(0) a b}R_{\rho \sigma}^{(0) c d}-16m^2aR_{\mu \nu}^{(0) a b}e_\rho{}^c e_\sigma{}^d+\\
    +64m^4a^2 e_\mu{}^a e_\nu{}^b e_\rho{}^c e_\sigma{}^d].
\end{aligned}
\end{equation}
Reading off the three pieces, we recognize a topological Gauss-Bonnet term, the Einstein-Hilbert (Palatini) action, and a cosmological-constant term, in that order; for negative $a$, this is General Relativity set in Anti-de Sitter space.

\subsubsection*{When $b_\mu{}^a = -\frac{1}{4} (R_\mu{}^a - \frac{1}{6} R e_\mu{}^a)$ - Weyl action}
Choosing instead the relation between $b$ and $e$ adopted in \cite{Kaku:1978nz, freedman_vanproeyen_2012} produces
\begin{equation}
    \begin{aligned}
             S_W=\frac{a_{CG}}{4}&\int d^4 x \epsilon^{\mu \nu \rho \sigma} \epsilon_{a b c d}
             \\&\left[R_{\mu \nu}^{(0) a b}+\frac{1}{2}\left(m e_\mu{}^{[a} R_\nu{}^{b]}-me_\nu{}^{[a} R_\mu{}^{b]}\right)-\frac{1}{3} m^2 R e_\mu{}^{[a} e_\nu{}^{b]}\right]\\
             &\left[R_{\rho \sigma}^{(0) c d}+\frac{1}{2}\left(m e_\rho{}^{[c} R_\sigma{}^{d]}-m e_\sigma{}^{[c} R_\rho{}^{d]}\right)-\frac{1}{3} m^2 R e_\rho{}^{[c} e_\sigma{}^{d]}\right].
    \end{aligned}
\end{equation}
After rescaling the vierbein via $\tilde{e}_\mu{}^{a}=m e_\mu{}^{a}$ and invoking the antisymmetry $R_{\mu \nu}^{(0)ab} = -R_{\nu \mu}^{(0)ab}$, each square bracket collapses to the Weyl conformal tensor $C_{\mu\nu}{}^{ab}$:
\begin{align}
             S_W=\frac{a_{CG}}{4}\int d^4 x \epsilon^{\mu \nu \rho \sigma} \epsilon_{a b c d}C_{\mu \nu}{}^{a b}C_{\rho \sigma}{}^{c d},
\end{align}
which is nothing but the well-known scale-invariant Weyl action in four dimensions, equally known in the form,
\begin{equation}
S_W =2a_{CG}\int \mathrm{d}^4 x\left(R_{\mu \nu} R^{\nu \mu}-\frac{1}{3} R^2\right).
\end{equation}

Being scale invariant, the Weyl action of WG contains no cosmological constant by construction, in either of the two forms derived above. Like CG, it is an attractive candidate for describing gravity at high energy scales (for recent developments along these lines, see \cite{Maldacena:2011mk, mannheim, Anastasiou:2016jix, ghilencea2023, Hell:2023rbf, Condeescu:2023izl}). Since WG itself arises from the SSB of CG, it is worth working out how a further SSB step, applied to WG, can in turn lead down to EG.

This is achieved by starting again from CG and introducing a scalar in the $\mathbf{15}$ of $SU(4)\sim SO(6)\sim SO(2,4)$, exactly as above. With the relation
\begin{equation}
\label{b-e_relation_repeat}
b_{\mu}{}^{a}=-\frac{1}{4}\left(R_{\mu}{}^{a}-\frac{1}{6}R\, e_{\mu}{}^{a}\right),
\end{equation}
between $b$ and $e$, the SSB of this $\mathbf{15}$-plet reproduces the Weyl action, as before. A second scalar is then added, this time in the second-rank antisymmetric representation of $SU(4)$, the $\mathbf{6}$; its spontaneous breaking, on top of the first, produces the Einstein-Hilbert action.

The two breakings are most easily tracked through the branching of the $\mathbf{15}$. Under $SU(2)\times SU(2)\times U(1)$, the subgroup left unbroken once $\langle\mathbf{15}\rangle$ acquires a vev,
\begin{align}
\label{SU4ToSU2SU2U1_repeat}
    SU(4)&\xrightarrow{\langle\mathbf{15}\rangle}SU(2)\times SU(2)\times U(1), \nonumber\\
    \mathbf{15}&=[(\mathbf{3},\mathbf{1})_0+(\mathbf{1},\mathbf{3})_0]+(\mathbf{1},\mathbf{1})_0 +(\mathbf{2},\mathbf{2})_{+2}+(\mathbf{2},\mathbf{2})_{-2}\,,
\end{align}
with $[(\mathbf{3},\mathbf{1})_0+(\mathbf{1},\mathbf{3})_0]$ the Lorentz generators $M_{ab}$, $(\mathbf{1},\mathbf{1})_0$ the dilatation generator $D$, and the two $(\mathbf{2},\mathbf{2})$ pieces the translation generators $P_a$ and conformal boost generators $K_a$ (up to a labeling convention between the two). $P_a$ and $K_a$ are thus broken already at this stage, while $M_{ab}$ and $D$ remain, matching the Lorentz and scale-invariant content of WG.

A second, complementary branching of the same $\mathbf{15}$ arises under the $SO(5)$ subgroup singled out by the $\mathbf{6}$-plet vev,
\begin{align}
\label{SU4ToSO5_repeat}
    SU(4)&\xrightarrow{\langle\mathbf{6}\rangle}SO(5), \nonumber\\
    \mathbf{15}&=\mathbf{10}+\mathbf{5}\,,
\end{align}
which under $SO(5)\supset SU(2)\times SU(2)$ further gives
\begin{align}
\label{SO5decomp}
    SO(5)& \supset SU(2) \times SU(2) \nonumber\\
    \mathbf{10}&=(\mathbf{3},\mathbf{1})+(\mathbf{1},\mathbf{3})
                 +(\mathbf{2},\mathbf{2})\,,\\
    \mathbf{5}&=(\mathbf{1},\mathbf{1})+(\mathbf{2},\mathbf{2})\,. \nonumber
\end{align}
The surviving $\mathbf{10}$ is made up of the Lorentz generators $M_{ab}$ together with the translation generators $P_a$; the broken $\mathbf{5}$ carries the dilatation singlet together with the $(\mathbf{2},\mathbf{2})$ associated with the conformal boosts $K_a$.

Together, the two steps read: $\langle\mathbf{15}\rangle$ breaks down spontaneously the generators $P_a$ and $K_a$, leaving $M_{ab}$ and $D$ unbroken, that is, WG; $\langle\mathbf{6}\rangle$, by giving a vev in its $(\mathbf{1},\mathbf{1})$ component under the decomposition \eqref{SO5decomp}, breaks spontaneously $D$ and contributes further to the breaking of the $K_a$ generators.

It was shown explicitly in \cite{Patellis:2025qbl} that the equivalence of diffeomorphisms and gauge transformations follows already from the on-shell vanishing of the torsion tensors $\tilde{R}_{\mu\nu}{}^{a}$ and $R_{\mu\nu}{}^{a}$ together with $F_{\mu\nu}=0$; the relation \eqref{b-e_relation_repeat} used above to obtain WG is not needed for this equivalence, and should instead be regarded as an additional, optional constraint singling out a particular class of solutions. The gauge-fixing used throughout this derivation, as elsewhere in the paper, is a standard technical device for exhibiting the physical degrees of freedom and the SSB pattern, and does not restrict the generality of the gauge-theoretic construction of gravity. This sequential $CG\to WG\to EG$ route, via a $\mathbf{15}$ followed by a $\mathbf{6}$, complements the two-vector-scalar mechanism of the next subsection, which reaches CG, WG, and EG directly, without an intermediate scale-invariant stage.

\subsection{Spontaneous symmetry breaking by introducing two scalars in vector representations}

The gauge group $SO(2,4)\sim SO(6)\sim SU(4)$ can also be broken using two scalars in the vector representation of $SO(6)$, instead of a single adjoint scalar. A first scalar in the $\mathbf{6}$, with a vev along $\langle \mathbf{1} \rangle$, breaks $SO(2,4)$ down to $Sp_4=SO(5)$, since
\begin{equation}
\begin{aligned}
    SO(6) &\supset SO(5) \\
    \mathbf{6} &=\mathbf{1}+\mathbf{5}.
\end{aligned}
\end{equation}
A second scalar in the $\mathbf{5}$ of $SO(5)$ then breaks the symmetry further, via
\begin{equation}
\begin{aligned}
    SO(5) &\supset SU(2)\times SU(2)\\
    \mathbf{5} &=(\mathbf{1},\mathbf{1})+(\mathbf{2},\mathbf{2}).
\end{aligned}
\end{equation}
As $SU(2)\times SU(2) \sim SO(4) \sim SO(1,3)$, giving this second $\mathbf{5}$ a vev along $\langle \mathbf{1},\mathbf{1}\rangle$ leaves precisely the Lorentz group intact. In terms of the $SO(2,4)$ algebra,
\begin{equation}
    \left[J_{A B}, J_{C D}\right] = \eta_{BC}J_{AD}+\eta_{AD} J_{BC} -\eta_{AC} J_{BD} -\eta_{BD} J_{AC},
\end{equation}
with $A,B=1,\dots,6$ and $\eta_{AB}=(-1,1,1,1,-1,1)$, a signature chosen precisely so that the two breaking steps act on a spacelike and a timelike direction respectively, the gauge field $A_\mu = \frac{1}{2}A_\mu{}^{AB}J_{AB}$ carries field strength
\begin{equation*}
    F_{\mu \nu}=[D_\mu , D_\nu]=\partial_\mu A_\nu -\partial_\nu A_\mu +[A_\mu ,A_\nu]\rightarrow
\end{equation*}
\begin{equation}
    F_{\mu \nu}{}^{AB}=\partial_\mu A_\nu{}^{AB} -\partial_\nu A_\mu{}^{AB} +A_\mu{}^{A}{}_{C}A_\nu{}^{CB}-A_\nu{}^{A}{}_{C}A_\mu{}^{CB}.
\end{equation}
An $SO(2,4)$-invariant quadratic action can then be built out of two vector scalars $\phi^E,\chi^F$ with mass scales $m_\phi \geq m_\chi$:
\begin{equation}
\begin{aligned}
    S_{SO(2,4)}=a_{CG}\int d^4x [&\epsilon^{\mu \nu \rho \sigma}\epsilon_{ABCDEF}\phi^E \chi^F m_\phi m_\chi \frac{1}{4} F_{\mu \nu}{}^{AB}F_{\rho \sigma}{}^{CD}+\\
    &+\lambda_\phi (\phi^E \phi_E -{m_\phi}^{-2})+\lambda_\chi (\chi^F \chi_F +{m_\chi}^{-2})],
\end{aligned}
\end{equation}
Setting $\phi^E=\phi^0=(0,0,0,0,0,{m_\phi}^{-1})$ as a gauge choice breaks the symmetry down to $SO(2,3)$,
\begin{equation}
    S_{SO(2,3)}=a_{CG}\int d^4x [\epsilon^{\mu \nu \rho \sigma}\epsilon_{ijklm} m_\chi \chi^m \frac{1}{4}F_{\mu \nu}{}^{ij}F_{\rho \sigma}{}^{kl} + \lambda_\chi (\chi^m\chi_m +{m_{\chi}^{-2}})],
\end{equation}
after which fixing the remaining scalar to $\chi^m=\chi^0=(0,0,0,0,{m_\chi}^{-1})$ removes the last four generators, leaving the six Lorentz generators $M_{ab}$ as the only unbroken symmetry. Parametrizing the surviving fields as
\begin{align}
       A_\mu{}^{j6}=m_\phi f_\mu{}^j \Rightarrow \begin{cases}
           A_\mu{}^{a6}={m_\phi}(b_\mu{}^a-e_\mu{}^a) \\
           A_\mu{}^{56}=-m_\phi \tilde{a}_\mu \\
       \end{cases}
       \text{and}\quad
     A_\mu{}^{a5}=-{m_\chi} (b_\mu{}^a+e_\mu{}^a),\qquad
    A_\mu{}^{ab}=\omega_\mu{}^{ab},
\end{align}
gives the field strength
\begin{align}\label{curv2}
F_{\mu \nu}{}^{a b}=&R_{\mu \nu}{}^{a b}+( {m_\chi}^2-{m_\phi}^2)\left(e_\mu{}^a e_\nu{}^b-e_\nu{}^a e_\mu{}^b+b_\mu{}^a b_\nu{}^b-b_\nu{}^a b_\mu{}^b\right)\nonumber\\
&-( {m_\chi}^2+{m_\phi}^2)(b_\mu{}^a e_\nu{}^b-b_\nu{}^a e_\mu{}^b+e_\mu{}^a b_\nu{}^b-e_\nu{}^a b_\mu{}^b),
\end{align}
together with
\begin{align}
F_{\mu \nu}{}^{a 5}&=-{m_\chi}[ T_{\mu \nu}{}^a(e)+T_{\mu \nu}{}^a(b)] +{m_\phi}^2[\tilde{a}_\mu( e_\nu{}^a -b_\nu{}^a)-\tilde{a}_\nu( e_\mu{}^a-b_\mu{}^a)].
\end{align}
As before, $F_{\mu\nu}{}^{a5}$ has no effect on the resulting quadratic action from \eqref{curv2}, so we set it to zero; this in turn forces $\tilde{a}_\mu=0$, and the theory comes out torsion-free \cite{Roumelioti:2024lvn}. The action then takes the form
\begin{equation}
\begin{aligned}\label{act2}
    S_{SO(1,3)} =\frac{a_{CG}}{4}\int &d^4x \epsilon^{\mu \nu \rho \sigma}\epsilon_{abcd}F_{\mu \nu}{ }^{a b} F_{\rho \sigma}{ }^{c d}= \\
    =\frac{a_{CG}}{4}\int & d^4x \epsilon^{\mu \nu \rho \sigma}\epsilon_{abcd}[R_{\mu \nu}{ }^{a b} R_{\rho \sigma}{ }^{c d} +\\
    &+2( {m_\chi}^2-{m_\phi}^2)R_{\mu \nu}{ }^{a b}(e_\rho{}^c e_\sigma{}^d-e_\sigma{}^c e_\rho{}^d+b_\rho{}^c b_\sigma{}^d-b_\sigma{}^c b_\rho{}^d)\\
    &-2( {m_\chi}^2+{m_\phi}^2)R_{\mu \nu}{ }^{a b}(b_\rho{}^c e_\sigma{}^d-b_\sigma{}^c e_\rho{}^d+e_\rho{}^c b_\sigma{}^d-e_\sigma{}^c b_\rho{}^d)\\
 &-2( {m_\chi}^4-{m_\phi}^4)\left(e_\mu{}^a e_\nu{}^b-e_\nu{}^a e_\mu{}^b+b_\mu{}^a b_\nu{}^b-b_\nu{}^a b_\mu{}^b\right) \\
&\qquad\qquad\qquad\qquad\times(b_\rho{}^c e_\sigma{}^d-b_\sigma{}^c e_\rho{}^d+e_\rho{}^c b_\sigma{}^d-e_\sigma{}^c b_\rho{}^d)\\
    &+( {m_\chi}^2-{m_\phi}^2)^2\left(e_\mu{}^a e_\nu{}^b-e_\nu{}^a e_\mu{}^b+b_\mu{}^a b_\nu{}^b-b_\nu{}^a b_\mu{}^b\right)\\
    &\qquad\qquad\qquad\qquad\times(e_\rho{}^c e_\sigma{}^d-e_\sigma{}^c e_\rho{}^d+b_\rho{}^c b_\sigma{}^d-b_\sigma{}^c b_\rho{}^d)\\
    & +( {m_\chi}^2+{m_\phi}^2)^2\left(b_\mu{}^a e_\nu{}^b-b_\nu{}^a e_\mu{}^b+e_\mu{}^a b_\nu{}^b-e_\nu{}^a b_\mu{}^b\right)\\
    &\qquad\qquad\qquad\qquad\times(b_\rho{}^c e_\sigma{}^d-b_\sigma{}^c e_\rho{}^d+e_\rho{}^c b_\sigma{}^d-e_\sigma{}^c b_\rho{}^d)].
\end{aligned}
\end{equation}
The component $F_{\mu\nu}{}^{56}$ is similarly absent from the action \cite{Roumelioti:2024lvn}, so
\begin{equation}
\begin{aligned}
    F_{\mu \nu}{}^{56}={m_\phi}[\partial_\mu \tilde{a}_\nu -\partial_\nu \tilde{a}_\mu-m_\chi(e_{\mu a}b_\nu{}^a-e_{\nu a}b_\mu{}^a)]=0
\end{aligned}
\end{equation}
can consistently be imposed; together with $\tilde{a}_\mu=0$, this once more constrains $e$ and $b$:
\begin{equation}\label{efa}
    e_{\mu a}b_\nu{}^a-e_{\nu a}b_\mu{}^a=0.
\end{equation}

\subsubsection*{When $b_\mu{}^a=ae_\mu{}^a$ - Einstein-Hilbert action in the presence of a cosmological constant}\label{casea}
This case proceeds exactly as before, yielding
\begin{equation}
\begin{aligned}
\label{34}
 S_{SO(1,3)} =\frac{a_{CG}}{4}\int d^4x &\epsilon^{\mu \nu \rho \sigma}\epsilon_{abcd}F_{\mu \nu}{ }^{a b} F_{\rho \sigma}{ }^{c d}= \\
    =\frac{a_{CG}}{4}\int d^4x \epsilon^{\mu \nu \rho \sigma}\epsilon_{abcd} & \Big[R_{\mu \nu}{ }^{a b} R_{\rho \sigma}{ }^{c d} +4\Big({m_\chi}^2(1-a)^2-{m_\phi}^2(1+a)^2\Big)R_{\mu \nu}{ }^{a b}e_\rho{}^c e_\sigma{}^d\\
    & +4\Big({m_\chi}^2(1-a)^2-{m_\phi}^2(1+a)^2\Big)^2 e_\mu{}^a e_\nu{}^b e_\rho{}^c e_\sigma{}^d\Big],
\end{aligned}
\end{equation}
i.e.\ again a Gauss-Bonnet term, an E-H (Palatini) term, and a cosmological-constant term, describing AdS space when ${m_\chi}^2/{m_\phi}^2>(1+a)^2/(1-a)^2$. For the equal-mass case $m_\phi=m_\chi\equiv m$,
\begin{equation}
\begin{aligned}
\label{35}
  S_{SO(1,3)} =\frac{a_{CG}}{4}\int & d^4x \epsilon^{\mu \nu \rho \sigma}\epsilon_{abcd}\Big[R_{\mu \nu}{ }^{a b} R_{\rho \sigma}{ }^{c d} -16{m}^2aR_{\mu \nu}{ }^{a b} e_\rho{}^c e_\sigma{}^d
    +64{m}^4 a^2e_\mu{}^a e_\nu{}^b e_\rho{}^c e_\sigma{}^d\Big],
\end{aligned}
\end{equation}
which for negative $a$ again places GR in AdS space, reproducing the single-adjoint-scalar result found earlier.

\subsubsection*{When $b_\mu{}^a=-\frac{1}{4}(R_\mu{}^a+\frac{1}{6}R e_\mu{}^a)$ and $m_\phi=m_\chi$ - Weyl action}
\label{caseb}

Taking instead this identification between $e$ and $b$, with $m \equiv m_\phi=m_\chi$, produces
\begin{equation}
    \begin{aligned}
             S=\frac{a_{CG}}{4}\int d^4 x \epsilon^{\mu \nu \rho \sigma} \epsilon_{a b c d}&\left[R_{\mu \nu}{}^{a b}+\frac{1}{2}\left(m e_\mu{}^{[a} R_\nu{}^{b]}-me_\nu{}^{[a} R_\mu{}^{b]}\right)-\frac{1}{3} m^2 R e_\mu{}^{[a} e_\nu{}^{b]}\right]\\
             &\left[R_{\rho \sigma}{}^{c d}+\frac{1}{2}\left(m e_\rho{}^{[c} R_\sigma{}^{d]}-m e_\sigma{}^{[c} R_\rho{}^{d]}\right)-\frac{1}{3} m^2 R e_\rho{}^{[c} e_\sigma{}^{d]}\right],
    \end{aligned}
\end{equation}
which, once $\tilde{e}_\mu{}^{a}=m e_\mu{}^{a}$ is substituted and the antisymmetry of $R_{\mu\nu}{}^{ab}$ is used, collapses once again to
\begin{equation}
             S=\frac{a_{CG}}{4}\int d^4 x \epsilon^{\mu \nu \rho \sigma} \epsilon_{a b c d}C_{\mu \nu}{}^{a b}C_{\rho \sigma}{}^{c d},
\end{equation}
namely the scale-invariant Weyl theory in four dimensions,
\begin{equation}\label{weyll}
S =2 a_{CG}\int \mathrm{d}^4 x\left(R_{\mu \nu} R^{\nu \mu}-\frac{1}{3} R^2\right).
\end{equation}

\section{Noncommutative (Fuzzy) Gravity}
\subsection{Gauge Theories on Noncommutative Spaces}

Gauge theories on noncommutative spaces are built using the notion of the covariant coordinate \cite{Madore:2000en}, which plays the role of the covariant derivative in the ordinary case. Consider a field $\phi(X_a)$ on a fuzzy space with noncommuting coordinates $X_a$, transforming under a gauge group $G$ as
\begin{equation}
\delta\phi(X) = \lambda(X)\phi(X)\,.
\end{equation}
A scalar $\lambda(X)$ corresponds to $G=U(1)$, while a Hermitian $P\times P$ matrix $\lambda(X)$ corresponds to $G=U(P)$. The coordinates themselves are naturally considered as gauge singlets,
\begin{equation}
\delta X_\alpha = 0\,,
\end{equation}
which means the naive product $X_a\phi$ fails to transform covariantly:
\begin{equation}
\delta(X_a \phi) = X_a \lambda(X) \phi \neq \lambda(X) X_a \phi\,.
\end{equation}
Just as in the commutative case, this is cured by defining a covariant coordinate $\mathcal{X}_a \equiv X_a + A_a$ satisfying
\begin{equation}
\delta(\mathcal{X}_a \phi) = \lambda \mathcal{X}_a \phi\,, \qquad
\delta(\mathcal{X}_a) = [\lambda, \mathcal{X}_a]\,,
\end{equation}
and $A_a$ transforms as
\begin{equation}
\delta A_a = -[X_a, \lambda] + [\lambda, A_a]\,,
\end{equation}
which is precisely what justifies interpreting $A_a$ as a gauge connection in this framework \cite{Aschieri:2005wm, Aschieri:2004vh}. Its field strength,
\begin{equation}
F_{ab} \equiv [X_a, A_b] - [X_b, A_a] + [A_a, A_b] - C_{ab}{}^c A_c = [\mathcal{X}_a, \phi_b] - C_{ab}{}^c \phi_c\,,
\end{equation}
transforms covariantly, $\delta F_{ab} = [\lambda, F_{ab}]$, and it is this object we use throughout the rest of the section to build gravity as a gauge theory on a covariant fuzzy space.

\subsection{The Background Space}
\label{sec3.2}

The background space we work with is four-dimensional de Sitter ($dS_4$) or anti-de Sitter ($AdS_4$) space, obtained by embedding into five-dimensional flat space via
\begin{equation}
\label{constrainteq}
    \eta_{\mu \nu}x^{\mu} x^{\nu} = s R^2,
\end{equation}
where $R$ sets the curvature radius, $\eta_{\mu\nu}=\mathrm{diag}(-1,1,1,1,s)$, and the sign $s=+1$ picks out $dS_4$, with isometry group $SO(1,4)$, while $s=-1$ gives $AdS_4$, with isometry $SO(2,3)$. Neither ten-generator isometry group is large enough to covariantize the five embedding coordinates and the local Lorentz symmetry at once, so we move up to their smallest covariant extensions: $SO(1,5)$ for $dS_4$, $SO(2,4)$ for $AdS_4$, each with the fifteen generators needed to accommodate both.

Pushing this idea further, in the spirit of Snyder and Yang as reviewed in App.~\ref{appA}, we enlarge the group once more, to the 21-generator $SO(1,6)$ (or $SO(2,5)$), giving the chain
\begin{align*}
SO(1,6) \supset SO(1,5) \supset SO(1,4) \supset SO(1,3)
\quad \\ \text{(or } SO(2,5) \supset SO(2,4) \supset SO(2,3) \supset SO(1,3)).
\end{align*}
Its 21 generators $J_{MN}$ ($M,N=0,\dots,6$) satisfy
\begin{equation}
[J_{MN},J_{RS}] = i\left(\eta_{MR}J_{NS} + \eta_{NS}J_{MR} - \eta_{NR}J_{MS} - \eta_{MS}J_{NR} \right),
\end{equation}
with $\eta_{MN} = \text{diag}(-1,1,1,1,s,1,1)$. Decomposing step by step down to $SO(1,3)$ yields the non-trivial commutators
\begin{equation}
\begin{gathered}
    \left[J_{ij},J_{kl}\right]=i\left(\eta_{i k}J_{j l}+\eta_{j l}J_{i k}-\eta_{j k}J_{i l}-\eta_{i l}J_{j k}\right),\\
    \left[J_{i j},J_{k6}\right]=i\left(\eta_{i k}J_{j6}-\eta_{j k}J_{i6}\right),\ \left[J_{i j},J_{k5}\right]=i\left(\eta_{i k}J_{j5}-\eta_{j k}J_{i5}\right),\\
    \left[J_{i j},J_{k4}\right]=i\left(\eta_{i k}J_{j4}-\eta_{j k}J_{i4}\right),\ \left[J_{i6},J_{j6}\right]=i J_{ij},\ \left[J_{i6},J_{j5}\right]=-i \eta_{ij}J_{56},\\
    \left[J_{i6},J_{j4}\right]=-i \eta_{ij}J_{46},\ \left[J_{i6},J_{56}\right]=iJ_{i5},\ \left[J_{i6},J_{46}\right]=iJ_{i4},\\
    \left[J_{i5},J_{j5}\right]=i J_{ij},\ \left[J_{i5},J_{j4}\right]=-i \eta_{ij}J_{45},\ \left[J_{i 5},J_{56}\right]=-i J_{i6},\\
    \left[J_{i 5},J_{45}\right]=i J_{i4},\ \left[J_{i4},J_{j4}\right]=i s J_{ij},\ \left[J_{i 4},J_{46}\right]=-i s J_{i6},\\
    \left[J_{i 4},J_{45}\right]=-i s J_{i5},\ \left[J_{56},J_{46}\right]=-i J_{45},\ \left[J_{56},J_{45}\right]=i J_{46},\ \left[J_{46},J_{45}\right]=-i s J_{56},
\end{gathered}
\end{equation}
where $i,j,k,l=0,\dots,3$ run over the four-dimensional indices and $\eta_{ij}=\mathrm{diag}(-1,1,1,1)$. Now identify, in the language of $SO(1,3)$,
\begin{equation}
    \Theta_{ij}=\hbar J_{ij}, \,  X_i=\lambda J_{i5},\  \text{and} \ P_i=\frac{\hbar}{\lambda}J_{i4}\, ,\qquad
    Q_i=\frac{\hbar}{\lambda}J_{i6}, \,  q= J_{56}, \,  p= J_{46}, \ \text{and} \ h=J_{45},
\end{equation}
under which the algebra becomes
\begin{equation}
\begin{gathered}
\label{backgroundAlgebra}
    [\Theta_{ij}, \Theta_{kl}] = i\hbar(\eta_{ik}\Theta_{jl} + \eta_{jl}\Theta_{ik} - \eta_{jk}\Theta_{il} - \eta_{il}\Theta_{jk}), \\
    [\Theta_{ij}, X_k] = \frac{i}{\hbar} (\eta_{ik}X_j - \eta_{jk}X_i), \
    [\Theta_{ij}, P_k] = \frac{i}{\hbar} (\eta_{ik}P_j - \eta_{jk}P_i), \\
    [\Theta_{ij}, Q_k] = \frac{i}{\hbar} (\eta_{ik}Q_j - \eta_{jk}Q_i), \  [Q_i, Q_j] = i\frac{\hbar}{\lambda^2} \Theta_{ij}, \
    [X_i, X_j] = i\frac{\lambda^2}{\hbar} \Theta_{ij}, \\
    [P_i, P_j] = i s \frac{\hbar}{\lambda^2} \Theta_{ij}, \
    [Q_i, X_j] = -i \frac{\hbar}{\lambda^2} \eta_{ij} q, \
    [Q_i, P_j] = -i \frac{\hbar^2}{\lambda^2} \eta_{ij} p, \\
    [X_i, P_j] = -i\hbar \eta_{ij} h, \
    [Q_i, q] = i \frac{\hbar}{\lambda^2} X_i, \
    [Q_i, p] = i P_i, \\
    [X_i, q] = -i \frac{\lambda^2}{\hbar} Q_i, \
    [X_i, h] = i \frac{\lambda^2}{\hbar} P_i, \\
    [P_i, p] = -i s Q_i, \
    [P_i, h] = -i s \frac{\hbar}{\lambda^2} X_i, \\
    [q, p] = -i h, \
    [q, h] = i p, \
    [p, h] = -i s q.
\end{gathered}
\end{equation}
This algebra thus exhibits Snyder-type noncommutativity of the coordinates, via $[X_i,X_j]=i\frac{\lambda^2}{\hbar}\Theta_{ij}$, Yang-type noncommutativity of the momenta, via $[P_i,P_j]=is\frac{\hbar}{\lambda^2}\Theta_{ij}$, a Heisenberg-like relation between them, via $[X_i,P_j]=-i\hbar\eta_{ij}h$, while it identifies the group that needs to be gauged.

\subsection{Gauge group and representation}
The natural symmetry to gauge is the isometry group of the background, $SO(1,4)$ for $dS_4$, $SO(2,3)$ for $AdS_4$. On a noncommutative space, however, both commutators and anticommutators of Lie-algebra elements enter the construction. Writing $\varepsilon(X)=\varepsilon^{a}(X) T_{a}$ and $\phi(X)=\phi^{a}(X) T_{a}$,
\begin{equation}
\label{anticom}
[\varepsilon, \phi]=\frac{1}{2}\{\varepsilon^{a}, \phi^{b}\}\left[T_{a}, T_{b}\right]+\frac{1}{2}\left[\varepsilon^{a}, \phi^{b}\right]\{T_{a}, T_{b}\},
\end{equation}
unlike the commutative case, the second term here does not vanish, so $\{T_a,T_b\}$ must also belong to the algebra. Rather than extend to the full (infinite-dimensional) universal enveloping algebra, it is enough to choose a representation for which enlarging the gauge group closes both operations at once: $SO(1,4)\to SO(1,5)\times U(1)$ for $dS_4$, $SO(2,3)\to SO(2,4)\times U(1)$ for $AdS_4$. This is exactly the structure already present in \eqref{backgroundAlgebra}; the extra $U(1)$ is required by the anticommutators and is unrelated to the earlier enlargement of the isometry group in Sec.~\ref{sec3.2}.

The two cases differ only in the sign of $s$, so it is enough to work through $SO(2,4)\times U(1)$. In the standard four-dimensional Dirac representation, where $\{\gamma_a,\gamma_b\}=-2\eta_{ab}\mathbb{1}_4$ ($a,b=1,\dots,4$), the generators read
\begin{equation}
    M_{ab} = -\frac{i}{4}[\gamma_a,\gamma_b],\quad
    P_a = -\frac{1}{2}\gamma_a(1 - \gamma_5),\quad
    K_a = \frac{1}{2}\gamma_a(1 + \gamma_5),\quad
    D = -\frac{1}{2}\gamma_5,\quad
    \mathbb{1}_4\ (U(1)),
\end{equation}
obeying the conformal-algebra commutators
\begin{equation}
\label{Lcom}
\begin{aligned}
[M_{ab}, M_{cd}] &= \eta_{bc} M_{ad} + \eta_{ad} M_{bc} - \eta_{ac} M_{bd} - \eta_{bd} M_{ac}, \\
[M_{ab}, P_c] &= \eta_{bc} P_a - \eta_{ac} P_b, \qquad
[M_{ab}, K_c] = \eta_{bc} K_a - \eta_{ac} K_b, \\
[P_a, D] &= P_a, \qquad [K_a, D] = -K_a, \qquad
[K_a, P_b] = -2(\eta_{ab} D + M_{ab}),
\end{aligned}
\end{equation}
and the accompanying anticommutators
\begin{equation}
\label{Lanticom}
\begin{aligned}
\{M_{ab}, M_{cd}\} &= \frac{1}{2}(\eta_{ac}\eta_{bd} - \eta_{bc} \eta_{ad}) - i \epsilon_{abcd} D, \qquad
\{M_{ab}, P_c\} = +i \epsilon_{abcd} P^d, \\
\{M_{ab}, K_c\} &= -i \epsilon_{abcd} K^d, \qquad
\{M_{ab}, D\} = 2M_{ab}D, \qquad
\{P_a, K_b\} = 4M_{ab}D + \eta_{ab}, \\
\{K_a, K_b\} &= \{P_a, P_b\} = -\eta_{ab}, \qquad
\{P_a, D\} = \{K_a, D\} = 0.
\end{aligned}
\end{equation}

\subsection{Fuzzy Gravity}

Four-dimensional noncommutative gravity on $AdS_4$ is realized as a gauge theory of $SO(2,4)\times U(1)$\footnote{The $dS_4$ case goes through in the same way.}. As in \cite{Manolakos_paper1}, define the covariant coordinate $\mathcal{X}_{\mu} \equiv X_{\mu} + A_{\mu}$, with connection
\begin{equation*}
    A_{\mu} = a_{\mu} \otimes \mathbb{1}_{4} + \omega_{\mu}{}^{ab} \otimes M_{ab} + e_{\mu}{}^{a} \otimes P_{a} + b_{\mu}{}^{a} \otimes K_{a} + \tilde{a}_{\mu} \otimes D,
\end{equation*}
whose covariant field strength is \cite{Manolakos_paper1, Madore_1992}
\begin{equation}
    \hat{F}_{\mu \nu} \equiv \left[\mathcal{X}_{\mu}, \mathcal{X}_{\nu}\right] - \kappa^2 \hat{\Theta}_{\mu \nu},
\end{equation}
with $\hat{\Theta}_{\mu \nu} \equiv \Theta_{\mu \nu} + \mathcal{B}_{\mu \nu}$, where the 2-form $\mathcal{B}_{\mu\nu}$ guarantees $\Theta$ transforms as it should. Expanded on the algebra,
\begin{equation}
    \hat{F}_{\mu \nu} = R_{\mu \nu} \otimes \mathbb{1}_{4} + \frac{1}{2} R_{\mu \nu}{}^{ab} \otimes M_{ab} + \tilde{R}_{\mu \nu}{}^{a} \otimes P_{a} + R_{\mu \nu}{}^{a} \otimes K_{a} + \tilde{R}_{\mu \nu} \otimes D\,.
\end{equation}

Breaking $SO(2,4)\times U(1)$ down to the Lorentz group calls for a scalar $\Phi(X)$ carrying the adjoint ($\mathbf{15}$-dimensional) representation of $SU(4)\sim SO(2,4)$, equivalently a rank-2 antisymmetric $SO(2,4)$ tensor, with a $U(1)$ charge so that the $U(1)$ is spontaneously broken along with the rest of the generators of $SO(2,4)$ apart from those that constitute the Lorentz group $SO(1,3)$ \cite{Manolakos_paper1, Manolakos_paper2, Roumelioti:2024lvn}. This gives the action
\begin{equation}
\mathcal{S} = \operatorname{Trtr} \left[\lambda \Phi(X) \epsilon^{\mu \nu \rho \sigma} \hat{F}_{\mu \nu} \hat{F}_{\rho \sigma} + \eta\left(\Phi(X)^2 - \lambda^{-2} \mathbb{1}_N \otimes \mathbb{1}_4\right)\right],
\end{equation}
with Lagrange multiplier $\eta$, mass scale $\lambda$, and
\begin{equation}
\Phi(X) = \phi(X) \otimes \mathbb{1}_4 + \phi^{ab}(X) \otimes M_{ab} + \tilde{\phi}^a(X) \otimes P_a + \phi^a(X) \otimes K_a + \tilde{\phi}(X) \otimes D.
\end{equation}
Fixing the gauge along the dilatation direction, following \cite{Manolakos_paper1, Manolakos_paper2, Roumelioti:2024lvn},
\begin{equation}
\Phi(X) = \left.\tilde{\phi}(X) \otimes D\right|_{\tilde{\phi} = -2 \lambda^{-1}} = -2 \lambda^{-1} \mathbb{1}_N \otimes D,
\end{equation}
and applying the (anti)commutators together with the on-shell traces reduces the action to
\begin{equation}
\mathcal{S}_{br} = \operatorname{Tr}\left(\frac{\sqrt{2}}{4} \varepsilon_{abcd} R_{mn}{}^{ab} R_{rs}{}^{cd} - 4 R_{mn} \tilde{R}_{rs}\right) \varepsilon^{mnrs},
\end{equation}
every other term, the Lagrange multiplier included, dropping out at this gauge choice.

What remains unbroken is an $SO(1,3)$ gauge symmetry. Taking the commutative limit and applying the appropriate dictionary between commutative and noncommutative fields turns this into the Palatini action, equivalently, the Einstein-Hilbert action with a cosmological constant \cite{Manolakos_paper2}, so ordinary GR with a cosmological constant is what one recovers at low energies.

\section{Unification of Conformal and Fuzzy Gravities with
Internal Interactions}
\label{sec:so216}
It was shown in \cite{Roumelioti:2024lvn} that $SO(10)$ internal interactions can be unified with CG by embedding both inside $SO(2,16)$, again exploiting the fact that the dimension of the tangent group of a manifold need not equal that of the manifold itself \cite{Patellis:2025qbl, Patellis:2024znm, Roumelioti:2025nku, roumelioti2407, Roumelioti:2025cxi, Roumelioti:2024lvn, Percacci:1984ai, Percacci_1991, Nesti_2008, Nesti_2010, Krasnov:2017epi, Chamseddine2010, Chamseddine2016, noncomtomos, Konitopoulos:2023wst, Weinberg:1984ke}. Since CG is obtained by gauging $SO(2,4) \sim SU(2,2) \sim SO(6) \sim SU(4)$ (the last two isomorphisms holding in Euclidean signature), the first step is to identify, starting from $SO(2,16)$, the centralizer $C_{SO(2,16)} (SO(2,4)) = SO(12)$, which is expected to break further to the internal symmetry group $SO(10)$.

For simplicity we work in Euclidean signature (the non-compact case is treated in \cite{Roumelioti:2024lvn}), starting from $SO(18)$ with fermions in the spinor $\mathbf{256}$, subject to the Weyl condition, and breaking to the maximal subgroup $SO(6)\times SO(12)$. The relevant representations decompose, following \cite{Slansky:1981yr, Feger_2020, Li:1973mq}, as
\begin{equation}\label{so18}
\begin{aligned}
SO(18) & \supset S O(6) \times S O(12) & & \\
\mathbf{18} & =(\mathbf{6},\mathbf{1}) + (\mathbf{1}, \mathbf{12}) & & \text { vector } \\
\mathbf{153} & =(\mathbf{15}, \mathbf{1}) + (\mathbf{6}, \mathbf{12}) + (\mathbf{1}, \mathbf{66}) & & \text { adjoint } \\
\mathbf{256} & =(\mathbf{4}, \overline{\mathbf{32}})+(\overline{\mathbf{4}}, \mathbf{32}) & & \text { spinor } \\
\mathbf{170} & =(\mathbf{1}, \mathbf{1})+(\mathbf{6}, \mathbf{12})+\left(\mathbf{20}^{\prime}, \mathbf{1}\right)+(\mathbf{1}, \mathbf{77}) & & \text { 2nd rank symmetric }
\end{aligned}
\end{equation}
This breaking is triggered by a vev along the $\langle\mathbf{1},\mathbf{1}\rangle$ component of a $\mathbf{170}$-plet scalar, and thus the fermions in the spinorial representation $\mathbf{256}$ of $SO(18)$ decompose as $(\mathbf{4},\overline{\mathbf{32}})+(\overline{\mathbf{4}},\mathbf{32})$ under the unbroken gauge group $SO(6)\times SO(12)$.

The breaking is completed by two further steps \cite{Roumelioti:2024lvn, Patellis:2025qbl}:
\begin{equation}
\begin{aligned}
&S O(6) \rightarrow S U(2) \times S U(2),
\end{aligned}
\end{equation}
driven, on the CG side, by a scalar in the antisymmetric $\mathbf{15}$ of $SO(6)$ (found inside the $\mathbf{153}$ of $SO(18)$ in Eq.~\eqref{so18} above), while on the internal side
\begin{equation}
\begin{aligned}
&S O(12) \rightarrow S O(10) \times[U(1)]_{\text{global}}
\end{aligned}
\end{equation}
follows from a scalar in the $\mathbf{77}$ (likewise found inside the $\mathbf{170}$ of $SO(18)$), whose branching under $SO(10)\times U(1)$ reads
\begin{equation}\label{seventyseven}
\mathbf{77}=(\mathbf{1})(4)+(\mathbf{1})(0)+(\mathbf{1})(-4)+(\mathbf{10})(2)+(\mathbf{10})(-2)+(\mathbf{54})(0),
\end{equation}
with the breaking itself triggered by a vev along the $(\mathbf{1})(4)$ component. $SO(12)$ could instead be broken using a scalar in the adjoint $\mathbf{66}$ (found inside the $\mathbf{153}$ of $SO(18)$), whose branching reads $\mathbf{66}=(\mathbf{1})(0)+(\mathbf{10})(2)+(\mathbf{10})(-2)+(\mathbf{45})(0)$; a vev along the $U(1)$-neutral $(\mathbf{1})(0)$ component would then leave $SO(10)\times U(1)$ as a genuine unbroken local gauge symmetry. Since the vev used here instead sits in the $(\mathbf{1})(4)$ component of the $\mathbf{77}$, which carries nonzero $U(1)$ charge, the corresponding generator is spontaneously broken and only the associated charge survives as a global symmetry, hence $SO(10)\times[U(1)]_\text{global}$.
In $SO(6)\times SO(12)$ language, then, these two breakings are driven by scalars in the $(\mathbf{15},\mathbf{1})$ and $(\mathbf{1},\mathbf{77})$, or in the $(\mathbf{15},\mathbf{1})$ and $(\mathbf{1},\mathbf{66})$. Following a single Weyl spinor in the $\mathbf{256}$ of $SO(18)$, after all relevant spontaneous breakings one is left with
\begin{equation}
    4\times \mathbf{16}_L(-1)
\end{equation}
under $SO(10)\times[U(1)]_\text{global}$. This four-fold multiplicity follows from group theory alone, since $SO(18)$ is of the chirality-preserving form $4n+2$. It could in principle be halved by a further Majorana condition, but this option is not available here: for an $SO(t,s)$ gauge theory (with $t$, $s$ the number of time- and space-like directions), Weyl and Majorana conditions can be imposed simultaneously only when $s-t\equiv 0 \pmod 8$ \cite{CHAPLINE1982461, polchinski_1998, D_Auria_2001, majoranaspinors, Ortín_2015}, and here $s-t=14$. Separating the four families into distinct flavours therefore remains an open problem, to which we return in the Conclusions.

From $SO(10)\times[U(1)]_\text{global}$ down to the SM, we follow the pattern set out in \cite{Patellis:2025qbl}: an intermediate stage through either the Pati-Salam group $SU(4)_C\times SU(2)_L\times SU(2)_R$ (with or without the left-right discrete symmetry $D$) or the minimal left-right model $SU(3)_C\times SU(2)_L\times SU(2)_R\times U(1)_{B-L}$ (again, with or without $D$), giving four cases in total: $422$, $422D$, $3221$, $3221D$. The routes down from $SO(10)$ use a $\mathbf{210}$ for $422$ and $3221D$, a $\mathbf{54}$ for $422$, and a $\mathbf{45}$ for $3221$; every case then reaches the SM through a $\mathbf{126}$-plet, with the electroweak Higgs housed in a $\mathbf{10}$. Writing $M_\text{GUT}$ for the scale where the three SM couplings meet and $M_I$ for the intermediate breaking scale, the four chains read:

\begin{description}
\item[$422$:] \(\displaystyle SO(10)\Big|_{M_\text{GUT}} \xrightarrow{\langle \mathbf{210}_H\rangle} SU(4)_C\times SU(2)_L\times SU(2)_R\Big|_{M_I} \xrightarrow{\langle\overline{\mathbf{126}}_H\rangle} \mathrm{SM}\)
\item[$422D$:] \(\displaystyle SO(10)\Big|_{M_\text{GUT}} \xrightarrow{\langle \mathbf{54}_H\rangle} SU(4)_C\times SU(2)_L\times SU(2)_R\times\mathcal{D}\Big|_{M_I} \xrightarrow{\langle\overline{\mathbf{126}}_H\rangle} \mathrm{SM}\)
\item[$3221$:] \(\displaystyle SO(10)\Big|_{M_\text{GUT}} \xrightarrow{\langle \mathbf{45}_H\rangle} SU(3)_C\times SU(2)_L\times SU(2)_R\times U(1)_{B-L}\Big|_{M_I} \xrightarrow{\langle\overline{\mathbf{126}}_H\rangle} \mathrm{SM}\)
\item[$3221D$:] \(\displaystyle SO(10)\Big|_{M_\text{GUT}} \xrightarrow{\langle \mathbf{210}_H\rangle} SU(3)_C\times SU(2)_L\times SU(2)_R\times U(1)_{B-L}\times\mathcal{D}\Big|_{M_I} \xrightarrow{\langle\overline{\mathbf{126}}_H\rangle} \mathrm{SM}\)
\end{description}
\label{chains}

Tracing representations through this chain requires the full set of branching rules,
\begin{equation}
\begin{aligned}
S O(12) & \supset S O(10) \times[U(1)]_{\text {global }} \\
\mathbf{12} & =(\mathbf{1})(2)+(\mathbf{1})(-2)+(\mathbf{10})(0) \\
\mathbf{66} & =(\mathbf{1})(0)+(\mathbf{10})(2)+(\mathbf{10})(-2)+(\mathbf{45})(0) \\
\mathbf{77} & =(\mathbf{1})(4)+(\mathbf{1})(0)+(\mathbf{1})(-4)+(\mathbf{10})(2)+(\mathbf{10})(-2)+(\mathbf{54})(0) \\
\mathbf{495} & =(\mathbf{45})(0)+(\mathbf{120})(2)+(\mathbf{120})(-2)+(\mathbf{210})(0) \\
\mathbf{792} & =(\mathbf{120})(0)+(\mathbf{126})(0)+(\overline{\mathbf{126}})(0)+(\mathbf{210})(2)+(\mathbf{210})(-2),
\end{aligned}
\end{equation}
where the Higgs $\mathbf{10}$ sits inside the $\mathbf{12}$ of $SO(12)$ and the $\overline{\mathbf{126}}$ needed for the intermediate step sits inside the $\mathbf{792}$; across the four scenarios, the $\mathbf{45}$ traces back to the $\mathbf{66}$, the $\mathbf{54}$ to the $\mathbf{77}$, and the $\mathbf{210}$ to the $\mathbf{792}$. One level up, in $SO(18)$,
\begin{equation}
\begin{aligned}
SO(18) \supset & SO(6) \times SO(12) \\
\mathbf{18}= & (\mathbf{6}, \mathbf{1})+(\mathbf{1}, \mathbf{12}) \\
\mathbf{3060}= & (\mathbf{15}, \mathbf{1})+(\mathbf{10}, \mathbf{12})+(\overline{\mathbf{10}}, \mathbf{12})+(\mathbf{15}, \mathbf{66})+(\mathbf{6}, \mathbf{220})+(\mathbf{1}, \mathbf{495}) \\
\mathbf{8568}= & (\mathbf{6}, \mathbf{1})+(\mathbf{15}, \mathbf{12})+(\mathbf{10}, \mathbf{66})+(\overline{\mathbf{10}}, \mathbf{66})+\\ & \qquad+(\mathbf{15}, \mathbf{220})+(\mathbf{6}, \mathbf{495})+ (\mathbf{1}, \mathbf{792}),
\end{aligned}
\end{equation}
which, cross-referenced against \eqref{so18}, shows where each piece comes from: the $\mathbf{12}$ descends from the $\mathbf{18}$, the $\mathbf{792}$ from the $\mathbf{8568}$, the $\mathbf{66}$ from the $\mathbf{153}$, the $\mathbf{495}$ from the $\mathbf{3060}$, and the $\mathbf{77}$ from the $\mathbf{170}$.

Extending this construction to Fuzzy Gravity introduces two additional requirements \cite{roumelioti2407}: the fermions must be chiral, to avoid Planck-scale masses, and, since FG is a matrix model, they must be organized in matrix (tensorial) representations. Both requirements are met by starting FG unification directly from $SO(6)\times SO(12)$, with fermions in $(\mathbf{4},\overline{\mathbf{32}})+(\overline{\mathbf{4}},\mathbf{32})$, a chiral, tensorial representation of the fuzzy space \cite{roumelioti2407, Chatzistavrakidis:2010xi, Chatzistavrakidis:2011toc}. Since the gauge-theoretic construction of FG yields the group $SO(6)\times U(1)\sim SO(2,4)\times U(1)$, the resulting scheme differs from the CG case above only by the extra $U(1)$ factor, which is identified, through the anticommutation relations of the construction, with the $U(1)$ appearing in $SO(12)\supset SO(10)\times U(1)$.

\subsection{Low-energy phenomenology}

The four breaking chains of the $SO(10)$ introduced above were examined at one loop in \cite{Patellis:2024znm}. Starting from the SM couplings at $M_Z$ \cite{ParticleDataGroup:2022pth} and running them up through each intermediate group, the matching conditions at $M_I$ and $M_\text{GUT}$ fix both scales for every chain, together with the unified coupling $g_{10}(M_\text{GUT})$. The intermediate scale ranges between $10^{10}$ and $10^{14}\,\text{GeV}$ depending on the chain, while $M_\text{GUT}$ lies between $10^{15}$ and $10^{16}\,\text{GeV}$, with $422$ and $3221$ giving the highest unification scales among the four.

Confronting these scales with proton-lifetime bounds, following the two-loop analysis of \cite{King:2021gnh}, shows that $422D$ and $3221D$ are already excluded by Super-Kamiokande, while $422$ and $3221$ remain viable; $3221$ would, however, also be excluded should Hyper-Kamiokande not observe proton decay during its planned exposure. Every breaking chain produces topological defects, monopoles at the GUT scale in all four cases, together with cosmic strings at the intermediate scale for $3221$ and $3221D$, since their intermediate group contains an abelian $U(1)_{B-L}$ factor. For $422$, $422D$, and $3221D$, however, inflation must occur after these defects form in order to avoid overclosing the Universe, which dilutes any resulting gravitational-wave signal beyond detectability. Only for $3221$ can inflation be placed strategically between the GUT-scale and intermediate-scale breakings, leaving its cosmic strings, and hence their gravitational-wave signal, undiluted and potentially observable; combined with the proton-lifetime bounds above, $3221$ is thus the only channel with both a surviving breaking scale and a detectable cosmic-string signal. The resulting string tension, evaluated at the intermediate scale, is
\begin{equation}
G\mu \simeq \frac{1}{2(\alpha_{2R}(M_I)+\alpha_1(M_I))}\frac{M_I^2}{M_\text{Pl}^2}\,,
\end{equation}
which, substituting the two-loop results of \cite{King:2021gnh}, gives $G\mu\simeq2.0\times10^{-17}$ \cite{Patellis:2025qbl}, comfortably below current bounds from LIGO, EPTA, NANOGrav, and PPTA, and within reach of upcoming gravitational-wave searches. The $422$ channel, by contrast, produces no cosmic strings at all, so it offers no analogous gravitational-wave probe.

\section{Conclusions}

The construction presented here rests on the observation, discussed in detail in \cite{Roumelioti:2024lvn} and previewed in our earlier proceedings contribution \cite{Roumelioti:2025nku}, that the tangent group of a curved manifold need not have the same dimension as the manifold itself. Together with the fact that gravity, like the internal interactions, can be formulated as a gauge theory, it allows all interactions to be unified through a single, larger tangent group acting entirely within four dimensions, without compactification or dimensional reduction. Gauging $SO(2,4)$ gives CG, whose SSB leads, depending on the vacuum chosen, to either Einstein or Weyl gravity at low energies; enlarging the tangent group to $SO(2,16)$ then incorporates $SO(10)$ internal interactions, with FG reaching the same endpoint starting from $SO(6)\times SO(12)$. Extra-dimensional model building thus remains a useful, though no longer strictly necessary, tool for unification, a point reinforced by the continued absence of direct experimental evidence for extra dimensions.

Any scheme of this kind must address how it evades the Coleman-Mandula (CM) theorem \cite{Coleman1967}, which relies crucially on full Poincar\'e invariance. In earlier constructions \cite{Percacci:1984ai, Percacci_1991, Nesti_2008, Nesti_2010, Krasnov:2017epi, Chamseddine2010, Chamseddine2016, Manolakos:2023hif, noncomtomos, Konitopoulos:2023wst}, the unifying group contained only the Lorentz subgroup of Poincar\'e, not the full group; here, by contrast, the full Poincar\'e group is present and is spontaneously broken, as shown in detail in \cite{Roumelioti:2024lvn}.

The low-energy phenomenology of the $SO(2,16)$ scheme, including the breaking-scale estimates, proton-lifetime constraints, and the cosmic-string gravitational-wave signal, is discussed in Sec.~\ref{sec:so216}.

Several directions remain open. Since the $SO(2,16)$ scheme cannot avoid a four-fold fermion degeneracy, as discussed above, it is natural to look for another unification group compatible with a simultaneous Weyl-Majorana condition. Requiring $s-t\equiv 0\pmod 8$, together with SSB chains reaching both $SO(1,3)$ and $SO(10)$ and a chiral ($4n+2$) structure, singles out $SO(1,17)$, with gravity gauged through $SO(1,5)$ rather than the conformal group, as the smallest group satisfying all three conditions. This is expected to roughly halve the fermion degeneracy remaining after the full SSB chain; the remainder would still need to be lifted by an appropriate choice of $SO(10)$-breaking scalars. A detailed account of this construction, together with the renormalization-group analysis of its breaking scales, will be presented elsewhere \cite{StefasZoupanos:inprep}.

A cosmological application also suggests itself. The successive SSB stages of the schemes discussed here leave behind a number of heavy particles with Planck-suppressed couplings to ordinary matter, which may be cosmologically stable and therefore constitute dark-matter candidates. We are investigating whether the second vielbein-like field of CG, the gauge field associated with the special-conformal generators, can play this role once the scalars responsible for the SSB are promoted from auxiliary to dynamical fields \cite{NandiStefasZoupanos:inprep}.

Finally, the higher-curvature terms present in these actions bring with them the Ostrogradski ghost problem common to higher-derivative gravity: a purely quadratic-curvature action in four dimensions is power-counting renormalizable \cite{Stelle:1977ry}, a property also exploited in the Starobinsky model of inflation \cite{Starobinsky:1980te}, but at the cost of a propagating, non-unitary spin-2 ghost. One proposed resolution imposes suitable boundary conditions \cite{Maldacena:2011mk, Hell:2023rbf, hell2023degrees}, though this has so far been demonstrated only at the classical, tree level. A related instability, the Boulware-Deser ghost of massive spin-2 fields \cite{Boulware:1972yc}, was later shown to be avoidable through an appropriate nonlinear potential \cite{deRham:2010kj, deRham:2010tw}, leading to ghost-free bimetric theory, in which a massless graviton is accompanied by a massive spin-2 partner (see \cite{SchmidtMay:2015vnx} for a review). It should also be noted though, that the Boulware-Deser ghost could be avoided in our Unification scheme based on $SO(2,16)$ by fixing its mass (via the SSB of the Conformal Gravity) at the Higuchi bound \cite{SchmidtMay:2015vnx}. Given the close relation between bimetric theory and the Weyl-symmetric actions studied here, these two instances of the ghost problem may be connected within the present framework; clarifying this connection is among the directions we intend to pursue.

\acknowledgments
We are grateful to Costas Bachas, Loriano Bonora, Thanassis Chatzistavrakidis, Jean-Pierre Derendinger, Vlado Dobrev, Zhenya Ivanov, Alex Kehagias, Tom Kephart, Spyros Konitopoulos, Vasilis Letsios, Dieter Lust, George Manolakos, Pantelis Manousselis, Carmelo Martin, Partha Nandi, Tomas Ortin, Gregory Patellis, Roberto Percacci, Danai Roumelioti, Manos Saridakis, Paul Sorba, Nicholas Tracas, Timo Weigand, and George Weiglein, for many helpful discussions throughout the development of the theories discussed in this work.

\appendix
\section{Brief Historical review}
\label{appA}
This appendix summarizes the constructions of Snyder \cite{Snyder:1946qz} and Yang \cite{yang1947}, which motivate the way the covariant space of Section \ref{sec3.2} is built.

Snyder \cite{Snyder:1946qz} proposed the first Lorentz-invariant, discretized model of spacetime, obtained by introducing a fundamental length scale and identifying spacetime coordinates with elements of the Lie algebra of the de Sitter group.

He worked with the four-dimensional de Sitter group $SO(1,4)$, whose generators satisfy
\begin{equation}
\left[J_{\mu \nu},J_{\rho \sigma}\right]=i\left(\eta_{\mu \rho}J_{\nu \sigma}+\eta_{\nu \sigma}J_{\mu \rho}-\eta_{\nu \rho}J_{\mu \sigma}-\eta_{\mu \sigma}J_{\nu \rho}\right)\, ,
\end{equation}
where $\mu,\nu,\rho,\sigma = 0, \dots, 4$, $J_{\mu \nu} = -J_{\nu \mu}$, and $\eta_{\mu \nu}$ is the five-dimensional Minkowski metric of signature $\operatorname{diag}(-,+,+,+,+)$.

Decomposing $SO(1,4)$ into its maximal subgroup $SO(1,3)$ produces three relations: \begin{equation}
\begin{gathered}
\left[J_{ij},J_{kl}\right]=i\left(\eta_{i k}J_{j l}+\eta_{j l}J_{i k}-\eta_{j k}J_{i l}-\eta_{i l}J_{j k}\right),\\ \left[J_{i j},J_{k4}\right]=i\left(\eta_{i k}J_{j4}-\eta_{j k}J_{i4}\right),\ \left[J_{i4},J_{j4}\right]=i J_{ij},\
\end{gathered}
\end{equation}
with $i,j,k,l = 0, \dots, 3$. Snyder connected these generators to physical quantities via
\begin{equation}
\label{identifications.X,TH}
\Theta_{ij} = \hbar J_{ij}, \quad X_i = \lambda J_{i4},
\end{equation}
$\lambda$ being the fundamental length scale, which gives
\begin{equation}
\begin{gathered} \left[\Theta_{ij},\Theta_{kl}\right] = i\hbar\left(\eta_{i k}\Theta_{j l}+\eta_{j l}\Theta_{i k}-\eta_{j k}\Theta_{i l}-\eta_{i l}\Theta_{j k}\right),\\ \left[\Theta_{ij},X_k\right] = i\hbar\left(\eta_{i k}X_j - \eta_{j k}X_i\right),\ \left[X_i,X_j\right] = \frac{i\lambda^2}{\hbar}\Theta_{ij},\
\end{gathered}
\end{equation}
where the noncommuting spacetime coordinates are now manifest.

Yang extended Snyder's construction \cite{yang1947} to incorporate continuous translations into a noncommutative spacetime, by considering the larger group $SO(1,5)$ \cite{Snyder:1946qz, kastrup_1966, Heckman_2015}, with generators satisfying
\begin{equation} \left[J_{mn},J_{rs}\right]=i\left(\eta_{mr}J_{ns}+\eta_{ns}J_{mr}-\eta_{nr}J_{ms}-\eta_{ms}J_{nr}\right),
\end{equation}
$m,n,r,s = 0, \dots, 5$, $\eta_{mn} = \operatorname{diag}(-1,1,1,1,1,1)$.
Decomposing $SO(1,5)$ down through $SO(1,4)$ to $SO(1,3)$ produces nine relations:
\begin{equation}
\begin{gathered}
\left[J_{ij},J_{kl}\right]=i\left(\eta_{i k}J_{j l}+\eta_{j l}J_{i k}-\eta_{j k}J_{i l}-\eta_{i l}J_{j k}\right),\ \left[J_{ij},J_{k5}\right]=i\left(\eta_{i k}J_{j5}-\eta_{j k}J_{i5}\right),\\ \left[J_{i5},J_{j5}\right]=i J_{ij},\ \left[J_{ij},J_{k4}\right]=i\left(\eta_{i k}J_{j4}-\eta_{j k}J_{i4}\right),\ \left[J_{i4},J_{j4}\right]=i J_{ij},\ \left[J_{i4},J_{j5}\right]=i \eta_{ij}J_{45},\\ \left[J_{ij},J_{45}\right]=0,\ \left[J_{i4},J_{45}\right]=-i J_{i5},\ \left[J_{i5},J_{45}\right]=i J_{i4},\
\end{gathered}
\end{equation}
Following the same logic as before, the generators are identified with physical quantities via
\begin{equation} \Theta_{ij} = \hbar J_{ij}, \quad X_i = \lambda J_{i5},
\end{equation}
with momenta defined as
\begin{equation} \label{identifications.P} P_i = \frac{\hbar}{\lambda} J_{i4}, \end{equation}
and $h=J_{45}$. This dictionary turns the algebra into
\begin{equation}
\begin{gathered}
\left[\Theta_{ij},\Theta_{kl}\right] = i \hbar\left(\eta_{i k}\Theta_{j l}+\eta_{j l}\Theta_{i k}-\eta_{j k}\Theta_{i l}-\eta_{i l}\Theta_{j k}\right),\ \left[\Theta_{ij},P_k\right] = i\hbar\left(\eta_{i k}P_j - \eta_{j k}P_i\right),\\ \left[P_i,P_j\right] = \frac{i \hbar}{\lambda^2} \Theta_{ij},\ \left[\Theta_{ij},X_k\right] = i\hbar\left(\eta_{i k}X_j - \eta_{j k}X_i\right),\ \left[X_i,X_j\right] = \frac{i\lambda^2}{\hbar} \Theta_{ij},\\ \left[X_i,P_j\right] = i\hbar \eta_{ij} h,\ \left[\Theta_{ij},h\right] = 0,\ \left[X_i,h\right] = \frac{i\lambda^2}{\hbar} P_i,\ \left[P_i,h\right] = -\frac{i\hbar}{\lambda^2} X_i.\
\end{gathered}
\end{equation}

Two consequences follow. First, since momenta now appear inside the Lie algebra together with the coordinates, they too fail to commute, so momentum space is quantized in the same sense as position space. Second, the coordinate-momentum commutator takes a Heisenberg-like form, as expected from quantum mechanics.

\bibliographystyle{JHEP}
\bibliography{bibliography}

\providecommand{\href}[2]{#2}\begingroup\raggedright\begin{thebibliography}{10}

\bibitem{Kaluza:1921}
T.~Kaluza, \emph{{Zum Unit\"{a}tsproblem der Physik}}, {\emph{Sitzungsber.
  Preuss. Akad. Wiss. Berlin (Math. Phys.)} (1921) 966}.

\bibitem{Klein:1926}
O.~Klein, \emph{{Quantentheorie und f\"{u}nfdimensionale
  Relativit\"{a}tstheorie}}, {\emph{Z. Phys.} {\bfseries 37} (1926) 895}.

\bibitem{Kerner:1968}
R.~Kerner, \emph{{Generalization of the {K}aluza-{K}lein Theory for an
  Arbitrary Non-Abelian Gauge Group}}, {\emph{Ann. Inst. H. Poincare Phys.
  Theor.} {\bfseries 9} (1968) 143}.

\bibitem{CHO1987358}
Y.~Cho, \emph{{Higher-Dimensional Unifications of Gravitation and Gauge
  Theories}}, \href{https://doi.org/10.1063/1.522434}{\emph{J. Math. Phys.}
  {\bfseries 16} (1975) 2029}.

\bibitem{Cho:1975sf}
Y.~M. Cho and P.~G.~O. Freund, \emph{{Nonabelian Gauge Fields in
  Nambu-Goldstone Fields}},
  \href{https://doi.org/10.1103/PhysRevD.12.1711}{\emph{Phys. Rev. D}
  {\bfseries 12} (1975) 1711}.

\bibitem{Witten:1983}
E.~Witten, \emph{{Fermion Quantum Numbers in {K}aluza-{K}lein Theory}},
  {\emph{Conf. Proc. C} {\bfseries 8306011} (1983) 227}.

\bibitem{Georgi1999}
H.~Georgi, \emph{{L}ie {A}lgebras {I}n {P}article {P}hysics: from {I}sospin To
  {U}nified {T}heories}.
\newblock Frontiers in Physics. {Westview Press}, 1999.

\bibitem{FRITZSCH1975193}
H.~Fritzsch and P.~Minkowski, \emph{{Unified Interactions of Leptons and
  Hadrons}}, {\emph{Ann. Phys.} {\bfseries 93} (1975) 193}.

\bibitem{CHAPLINE1982461}
G.~Chapline and R.~Slansky, \emph{{Dimensional Reduction and Flavor
  Chirality}}, {\emph{Nucl. Phys. B} {\bfseries 209} (1982) 461}.

\bibitem{Green2012-ul}
M.~Green, J.~Schwarz and E.~Witten, \emph{Superstring Theory}, vol.~1 \& 2.
\newblock {Cambridge University Press}, 1988.

\bibitem{polchinski_1998}
J.~Polchinski, \emph{{String theory}}, vol.~{1 \& 2}.
\newblock Cambridge University Press, 1998.

\bibitem{Lust:1989tj}
D.~Lust and S.~Theisen, \emph{{Lectures on {S}tring {T}heory}}, vol.~346.
\newblock Springer, 1989.

\bibitem{GROSS1985253}
D.~Gross, J.~Harvey, E.~Martinec and R.~Rohm, \emph{{Heterotic String Theory.
  1. The Free Heterotic String}},
  \href{https://doi.org/10.1016/0550-3213(85)90394-3}{\emph{Nucl. Phys. B}
  {\bfseries 256} (1985) 253}.

\bibitem{forgacs}
P.~Forg{\aa}cs and N.~Manton, \emph{{Space-Time Symmetries in Gauge Theories}},
  {\emph{Commun. Math. Phys.} {\bfseries 72} (1980) 15}.

\bibitem{MANTON1981502}
N.~Manton, \emph{{Fermions and Parity Violation in Dimensional Reduction
  Schemes}}, {\emph{Nucl. Phys. B} {\bfseries 193} (1981) 502}.

\bibitem{Kubyshin:1989vd}
Y.~A. Kubyshin, I.~P. Volobuev, J.~M. Mourao and G.~Rudolph, \emph{{Dimensional
  Reduction of Gauge Theories, Spontaneous Compactification and Model
  Building}}, vol.~349.
\newblock Springer, 1989.

\bibitem{KAPETANAKIS19924}
D.~Kapetanakis and G.~Zoupanos, \emph{{Coset Space Dimensional Reduction of
  Gauge Theories}}, {\emph{Phys. Rept.} {\bfseries 219} (1992) 4}.

\bibitem{LUST1985309}
D.~Lust and G.~Zoupanos, \emph{{Dimensional Reduction of {T}en-dimensional
  $E_8$ Gauge Theory Over a Compact Coset Space}},
  \href{https://doi.org/10.1016/0370-2693(85)91236-5}{\emph{Phys. Lett. B}
  {\bfseries 165} (1985) 309}.

\bibitem{SCHERK197961}
J.~Scherk and J.~Schwarz, \emph{{How to Get Masses from Extra Dimensions}},
  {\emph{Nucl. Phys. B} {\bfseries 153} (1979) 61}.

\bibitem{Manousselis_2004}
P.~Manousselis and G.~Zoupanos, \emph{{Dimensional Reduction of
  {T}en-Dimensional Supersymmetric Gauge Theories in the $N=1$, $D=4$
  Superfield Formalism}},
  \href{https://doi.org/10.1088/1126-6708/2004/11/025}{\emph{JHEP} {\bfseries
  2004} (2004) 025}, [\href{https://arxiv.org/abs/hep-ph/0406207}{{\ttfamily
  hep-ph/0406207}}].

\bibitem{Chatzistavrakidis:2009mh}
A.~Chatzistavrakidis and G.~Zoupanos, \emph{{Dimensional Reduction of the
  Heterotic String over Nearly-{K}\"{a}hler Manifolds}},
  \href{https://doi.org/10.1088/1126-6708/2009/09/077}{\emph{JHEP} {\bfseries
  09} (2009) 077}, [\href{https://arxiv.org/abs/0905.2398}{{\ttfamily
  0905.2398}}].

\bibitem{Irges:2011de}
N.~Irges and G.~Zoupanos, \emph{{Reduction of $N=1$, $E_8$ {SYM} over
  $SU(3)/U(1)\times U(1)\times\mathbb{Z}_3$ and Its Four-Dimensional Effective
  Action}}, \href{https://doi.org/10.1016/j.physletb.2011.03.005}{\emph{Phys.
  Lett. B} {\bfseries 698} (2011) 146},
  [\href{https://arxiv.org/abs/1102.2220}{{\ttfamily 1102.2220}}].

\bibitem{Manolakos:2020cco}
G.~Manolakos, G.~Patellis and G.~Zoupanos, \emph{{$N=1$ Trinification from
  Dimensional Reduction of $N=1$, 10{D} $E_8$ over $SU(3)/U(1)\times
  U(1)\times\mathbb{Z}_3$}},
  \href{https://doi.org/10.1016/j.physletb.2020.136031}{\emph{Phys. Lett. B}
  {\bfseries 813} (2021) 136031},
  [\href{https://arxiv.org/abs/2009.07059}{{\ttfamily 2009.07059}}].

\bibitem{Patellis:2024dfl}
G.~Patellis, W.~Porod and G.~Zoupanos, \emph{{Split NMSSM from dimensional
  reduction of a $10D$, $\mathcal{N}=1$, $E_8$ theory over a modified flag
  manifold}}, \href{https://doi.org/10.1007/JHEP01(2024)021}{\emph{JHEP}
  {\bfseries 1} (2024) 021},
  [\href{https://arxiv.org/abs/2307.10014}{{\ttfamily 2307.10014}}].

\bibitem{utiyama}
R.~Utiyama, \emph{Invariant theoretical interpretation of interaction},
  \href{https://doi.org/10.1103/PhysRev.101.1597}{\emph{Phys. Rev.} {\bfseries
  101} (1956) 1597}.

\bibitem{kibble1961}
T.~W.~B. Kibble, \emph{{Lorentz Invariance and the Gravitational Field}},
  \href{https://doi.org/10.1063/1.1703702}{\emph{J. Math. Phys.} (1961)
  212--221}.

\bibitem{Sciama}
D.~Sciama, \emph{{Recent developments in the theory of gravitational
  radiation}}, \href{https://doi.org/10.1007/BF00755934}{\emph{Gen. Rel. Grav.}
  {\bfseries 3} (1972) 149--165}.

\bibitem{Umezawa}
H.~Umezawa, \emph{{Dynamical rearrangement of symmetries}},
  \href{https://doi.org/10.1007/BF02721035}{\emph{Nuovo Cim. A} {\bfseries 40}
  (1965) 450--475}.

\bibitem{Matsumoto}
H.~Matsumoto, N.~J. Papastamatiou and H.~Umezawa, \emph{Infrared effect of the
  goldstone boson and the order parameter},
  \href{https://doi.org/10.1103/PhysRevD.12.1836}{\emph{Phys. Rev. D}
  {\bfseries 12} (1975) 1836--1839}.

\bibitem{macdowell}
S.~W. MacDowell and F.~Mansouri, \emph{Unified geometric theory of gravity and
  supergravity}, \href{https://doi.org/10.1103/PhysRevLett.38.739}{\emph{Phys.
  Rev. Lett.} {\bfseries 38} (1977) 739}.

\bibitem{Ivanov:1980tw}
E.~Ivanov and J.~Niederle, \emph{{On Gauge Formulations of Gravitation
  Theories}},  in \emph{{9th International Colloquium on Group Theoretical
  Methods in Physics}}, 1980.

\bibitem{Ivanov:1981wn}
E.~Ivanov and J.~Niederle, \emph{{Gauge Formulation of Gravitation Theories. 1.
  The Poincare, De Sitter and Conformal Cases}},
  \href{https://doi.org/10.1103/PhysRevD.25.976}{\emph{Phys. Rev. D} {\bfseries
  25} (1982) 976}.

\bibitem{stellewest}
K.~S. Stelle and P.~C. West, \emph{Spontaneously broken de sitter symmetry and
  the gravitational holonomy group},
  \href{https://doi.org/10.1103/PhysRevD.21.1466}{\emph{Phys. Rev. D}
  {\bfseries 21} (1980) 1466}.

\bibitem{Kibble:1985sn}
T.~Kibble and K.~Stelle, \emph{{Gauge theories of gravity and supergravity}},
  {\emph{Progress in Quantum Field Theory} (1985) }.

\bibitem{freedman_vanproeyen_2012}
D.~Z. Freedman and A.~Van~Proeyen, \emph{{Supergravity}}.
\newblock Cambridge Univ. Press, 2012,
  \href{https://doi.org/10.1017/CBO9781139026833}{10.1017/CBO9781139026833}.

\bibitem{Ortín_2015}
T.~Ortín, \emph{Gravity and Strings}.
\newblock Cambridge Monographs on Mathematical Physics. {Cambridge University
  Press}, 2015,
  \href{https://doi.org/10.1017/CBO9781139019750}{10.1017/CBO9781139019750}.

\bibitem{castellani}
L.~Castellani, \emph{$os_p(1|4)$ supergravity and its noncommutative
  extension}, \href{https://doi.org/10.1103/physrevd.88.025022}{\emph{Phys.
  Rev. D} {\bfseries 88} (2013) }.

\bibitem{Chatzistavrakidis_2018}
A.~Chatzistavrakidis, L.~Jonke, D.~Jurman, G.~Manolakos, P.~Manousselis and
  G.~Zoupanos, \emph{{Noncommutative Gauge Theory and Gravity in Three
  Dimensions}}, \href{https://doi.org/10.1002/prop.201800047}{\emph{Fortsch.
  Phys.} {\bfseries 66} (2018) 1800047},
  [\href{https://arxiv.org/abs/1802.07550}{{\ttfamily 1802.07550}}].

\bibitem{Manolakos_paper1}
G.~Manolakos, P.~Manousselis and G.~Zoupanos, \emph{{Four-dimensional Gravity
  on a Covariant Noncommutative Space}},
  \href{https://doi.org/10.1007/JHEP08(2020)001}{\emph{JHEP} {\bfseries 08}
  (2020) 001}, [\href{https://arxiv.org/abs/1902.10922}{{\ttfamily
  1902.10922}}].

\bibitem{manolakosphd}
G.~Manolakos, \emph{{Construction of gravitational models as noncommutative
  gauge theories}}, Ph.D. thesis, Natl. Tech. U., Athens, 2019.

\bibitem{Manolakos_paper2}
G.~Manolakos, P.~Manousselis and G.~Zoupanos, \emph{Four-dimensional gravity on
  a covariant noncommutative space ({II})},
  \href{https://doi.org/10.1002/prop.202100085}{\emph{Fortsch. Phys.}
  {\bfseries 69} (2021) 2100085}.

\bibitem{Manolakos:2022universe}
G.~Manolakos, P.~Manousselis, D.~Roumelioti, S.~Stefas and G.~Zoupanos,
  \emph{{A Matrix Model of Four-Dimensional Noncommutative Gravity}},
  \href{https://doi.org/10.3390/universe8040215}{\emph{Universe} {\bfseries 8}
  (2022) 215}.

\bibitem{Manolakos:2023hif}
G.~Manolakos, P.~Manousselis, D.~Roumelioti, S.~Stefas and G.~Zoupanos,
  \emph{{Intertwining noncommutativity with gravity and particle physics}},
  \href{https://doi.org/10.1140/epjs/s11734-023-00830-8}{\emph{Eur. Phys. J.
  ST} {\bfseries 232} (2023) 3607},
  [\href{https://arxiv.org/abs/2305.11785}{{\ttfamily 2305.11785}}].

\bibitem{roumelioti2407}
D.~Roumelioti, S.~Stefas and G.~Zoupanos, \emph{{Fuzzy Gravity:
  Four-Dimensional Gravity on a Covariant Noncommutative Space and Unification
  with Internal Interactions}},
  \href{https://doi.org/10.1002/prop.202400126}{\emph{Fortsch. Phys.}
  {\bfseries 72} (2024) 2400126},
  [\href{https://arxiv.org/abs/2407.07044}{{\ttfamily 2407.07044}}].

\bibitem{weyl}
H.~Weyl, \emph{Gravitation and the electron},
  \href{https://doi.org/10.1073/pnas.15.4.323}{\emph{Proc. Nat. Acad. Sci.}
  {\bfseries 15} (1929) 323--334}.

\bibitem{weyl1929}
H.~Weyl, \emph{{Elektron und Gravitation}},
  \href{https://doi.org/10.1007/BF01339504}{\emph{Z. Phys.} {\bfseries 56}
  (1929) 330--352}.

\bibitem{Addazi:2024xkg}
A.~Addazi, S.~Capozziello, A.~Marciano and G.~Meluccio, \emph{{Gravity from
  Pre-geometry}}, \href{https://doi.org/10.1088/1361-6382/ada767}{\emph{Class.
  Quant. Grav.} {\bfseries 42} (2025) 045012},
  [\href{https://arxiv.org/abs/2409.02200}{{\ttfamily 2409.02200}}].

\bibitem{KAKU1977304}
M.~Kaku, P.~Townsend and P.~{Van Nieuwenhuizen}, \emph{Gauge theory of the
  conformal and superconformal group},
  \href{https://doi.org/10.1016/0370-2693(77)90552-4}{\emph{Phys. Lett. B}
  {\bfseries 69} (1977) 304--308}.

\bibitem{Roumelioti:2024lvn}
D.~Roumelioti, S.~Stefas and G.~Zoupanos, \emph{{Unification of conformal
  gravity and internal interactions}},
  \href{https://doi.org/10.1140/epjc/s10052-024-12949-6}{\emph{Eur. Phys. J. C}
  {\bfseries 84} (2024) 577},
  [\href{https://arxiv.org/abs/2403.17511}{{\ttfamily 2403.17511}}].

\bibitem{Percacci:1984ai}
R.~Percacci, \emph{{Spontaneous Soldering}},
  \href{https://doi.org/10.1016/0370-2693(84)90171-0}{\emph{Phys. Lett. B}
  {\bfseries 144} (1984) 37}.

\bibitem{Percacci_1991}
R.~Percacci, \emph{{The Higgs phenomenon in quantum gravity}},
  \href{https://doi.org/10.1016/0550-3213(91)90510-5}{\emph{Nucl. Phys. B}
  {\bfseries 353} (1991) 271},
  [\href{https://arxiv.org/abs/0712.3545}{{\ttfamily 0712.3545}}].

\bibitem{Nesti_2008}
F.~Nesti and R.~Percacci, \emph{{Graviweak Unification}},
  \href{https://doi.org/10.1088/1751-8113/41/7/075405}{\emph{J. Phys. A}
  {\bfseries 41} (2008) 075405},
  [\href{https://arxiv.org/abs/0706.3307}{{\ttfamily 0706.3307}}].

\bibitem{Nesti_2010}
F.~Nesti and R.~Percacci, \emph{{Chirality in unified theories of gravity}},
  \href{https://doi.org/10.1103/PhysRevD.81.025010}{\emph{Phys. Rev. D}
  {\bfseries 81} (2010) 025010},
  [\href{https://arxiv.org/abs/0909.4537}{{\ttfamily 0909.4537}}].

\bibitem{Chamseddine2010}
A.~H. Chamseddine and V.~Mukhanov, \emph{{{G}ravity with de {S}itter and
  {U}nitary {T}angent {G}roups}},
  \href{https://doi.org/10.1007/JHEP03(2010)033}{\emph{JHEP} {\bfseries 03}
  (2010) 033}, [\href{https://arxiv.org/abs/1002.0541}{{\ttfamily 1002.0541}}].

\bibitem{Chamseddine2016}
A.~H. Chamseddine and V.~Mukhanov, \emph{{On {U}nification of {G}ravity and
  {G}auge {I}nteractions}},
  \href{https://doi.org/10.1007/JHEP03(2016)020}{\emph{JHEP} {\bfseries 03}
  (2016) 020}, [\href{https://arxiv.org/abs/1602.02295}{{\ttfamily
  1602.02295}}].

\bibitem{Krasnov:2017epi}
K.~Krasnov and R.~Percacci, \emph{{Gravity and Unification: A review}},
  \href{https://doi.org/10.1088/1361-6382/aac58d}{\emph{Class. Quant. Grav.}
  {\bfseries 35} (2018) 143001},
  [\href{https://arxiv.org/abs/1712.03061}{{\ttfamily 1712.03061}}].

\bibitem{Konitopoulos:2023wst}
S.~Konitopoulos, D.~Roumelioti and G.~Zoupanos, \emph{{Unification of Gravity
  and Internal Interactions}},
  \href{https://doi.org/10.1002/prop.202300226}{\emph{Fortsch. Phys.}
  {\bfseries 2023} (2023) 2300226},
  [\href{https://arxiv.org/abs/2309.15892}{{\ttfamily 2309.15892}}].

\bibitem{noncomtomos}
P.~Schupp, K.~Anagnostopoulos and G.~Zoupanos, \emph{Noncommutativity and
  physics}, \href{https://doi.org/10.1140/epjs/s11734-024-01086-6}{\emph{Eur.
  Phys. J. ST} {\bfseries 232} (2024) 1}.

\bibitem{Patellis:2024znm}
G.~Patellis, N.~Tracas and G.~Zoupanos, \emph{{From the unification of
  conformal and fuzzy gravities with internal interactions to the SO(10) GUT
  and the particle physics standard model}},
  \href{https://doi.org/10.1140/epjc/s10052-025-13938-z}{\emph{Eur. Phys. J. C}
  {\bfseries 85} (2025) 328},
  [\href{https://arxiv.org/abs/2412.02786}{{\ttfamily 2412.02786}}].

\bibitem{Roumelioti:2025cxi}
D.~Roumelioti, S.~Stefas and G.~Zoupanos, \emph{{Unification of Conformal and
  Fuzzy Gravities with Internal Interactions based on the SO(10) GUT}},  in
  \emph{{11th Mathematical Physics Meeting}}, 3, 2025,
  \href{https://arxiv.org/abs/2503.04574}{{\ttfamily 2503.04574}}.

\bibitem{Patellis:2025qbl}
G.~Patellis, D.~Roumelioti, S.~Stefas and G.~Zoupanos, \emph{{Unification of
  Conformal and Fuzzy Gravities With Internal Interactions Resulting in SO(10)
  and a Possible Probe Through Stochastic Gravitational Wave Background}},
  \href{https://doi.org/10.1002/prop.70014}{\emph{Fortsch. Phys.} {\bfseries
  73} (2025) e70014}, [\href{https://arxiv.org/abs/2504.06142}{{\ttfamily
  2504.06142}}].

\bibitem{Roumelioti:2025nku}
D.~Roumelioti, S.~Stefas and G.~Zoupanos, \emph{{Conformal and Fuzzy Gravities
  and their Unification with Internal Interactions}},
  \href{https://doi.org/10.22323/1.490.0153}{\emph{PoS} {\bfseries CORFU2024}
  (2025) 153}.

\bibitem{Kaku:1978nz}
M.~Kaku, P.~K. Townsend and P.~van Nieuwenhuizen, \emph{Properties of conformal
  supergravity}, \href{https://doi.org/10.1103/PhysRevD.17.3179}{\emph{Phys.
  Rev. D} {\bfseries 17} (Jun, 1978) 3179--3187}.

\bibitem{Li:1973mq}
L.-F. Li, \emph{{Group Theory of the Spontaneously Broken Gauge Symmetries}},
  \href{https://doi.org/10.1103/PhysRevD.9.1723}{\emph{Phys. Rev. D} {\bfseries
  9} (1974) 1723}.

\bibitem{Chamseddine:2002fd}
A.~H. Chamseddine, \emph{{Invariant actions for noncommutative gravity}},
  \href{https://doi.org/10.1063/1.1572199}{\emph{J. Math. Phys.} {\bfseries 44}
  (2003) 2534}, [\href{https://arxiv.org/abs/hep-th/0202137}{{\ttfamily
  hep-th/0202137}}].

\bibitem{Maldacena:2011mk}
J.~Maldacena, \emph{{Einstein Gravity from Conformal Gravity}},
  \href{https://arxiv.org/abs/1105.5632}{{\ttfamily 1105.5632}}.

\bibitem{mannheim}
P.~Mannheim, \emph{{Making the Case for Conformal Gravity}},
  \href{https://doi.org/10.1007/s10701-011-9608-6}{\emph{Found. Phys.}
  {\bfseries 42} (2012) 388--420}.

\bibitem{Anastasiou:2016jix}
G.~Anastasiou and R.~Olea, \emph{{From conformal to Einstein Gravity}},
  \href{https://doi.org/10.1103/PhysRevD.94.086008}{\emph{Phys. Rev. D}
  {\bfseries 94} (2016) 086008},
  [\href{https://arxiv.org/abs/1608.07826}{{\ttfamily 1608.07826}}].

\bibitem{ghilencea2023}
D.~M. Ghilencea, \emph{{Weyl conformal geometry vs Weyl anomaly}},
  \href{https://doi.org/10.1007/JHEP10(2023)113}{\emph{JHEP} {\bfseries 10}
  (2023) 113}, [\href{https://arxiv.org/abs/2309.11372}{{\ttfamily
  2309.11372}}].

\bibitem{Hell:2023rbf}
A.~Hell, D.~Lust and G.~Zoupanos, \emph{{On the ghost problem of conformal
  gravity}}, \href{https://doi.org/10.1007/JHEP08(2023)168}{\emph{JHEP}
  {\bfseries 08} (2023) 168},
  [\href{https://arxiv.org/abs/2306.13714}{{\ttfamily 2306.13714}}].

\bibitem{Condeescu:2023izl}
C.~Condeescu, D.~M. Ghilencea and A.~Micu, \emph{{Weyl quadratic gravity as a
  gauge theory and non-metricity vs torsion duality}},
  \href{https://doi.org/10.1140/epjc/s10052-024-12644-6}{\emph{Eur. Phys. J. C}
  {\bfseries 84} (2024) 292},
  [\href{https://arxiv.org/abs/2312.13384}{{\ttfamily 2312.13384}}].

\bibitem{Madore:2000en}
J.~Madore, S.~Schraml, P.~Schupp and J.~Wess, \emph{{Gauge theory on
  noncommutative spaces}},
  \href{https://doi.org/10.1007/s100520050012}{\emph{Eur. Phys. J. C}
  {\bfseries 16} (2000) 161},
  [\href{https://arxiv.org/abs/hep-th/0001203}{{\ttfamily hep-th/0001203}}].

\bibitem{Aschieri:2005wm}
P.~Aschieri, J.~Madore, P.~Manousselis and G.~Zoupanos, \emph{{Renormalizable
  theories from fuzzy higher dimensions}},  in \emph{{3rd Summer School in
  Modern Mathematical Physics}}, pp.~135--145, 3, 2005,
  \href{https://arxiv.org/abs/hep-th/0503039}{{\ttfamily hep-th/0503039}}.

\bibitem{Aschieri:2004vh}
P.~Aschieri, J.~Madore, P.~Manousselis and G.~Zoupanos, \emph{{Unified theories
  from fuzzy extra dimensions}},
  \href{https://doi.org/10.1002/prop.200410168}{\emph{Fortsch. Phys.}
  {\bfseries 52} (2004) 718--723},
  [\href{https://arxiv.org/abs/hep-th/0401200}{{\ttfamily hep-th/0401200}}].

\bibitem{Madore_1992}
J.~Madore, \emph{The fuzzy sphere},
  \href{https://doi.org/10.1088/0264-9381/9/1/008}{\emph{Class. Quant. Grav.}
  {\bfseries 9} (1992) 69}.

\bibitem{Weinberg:1984ke}
S.~Weinberg, \emph{{Generalized Theories of Gravity and Supergravity in Higher
  Dimensions}}, {\emph{{Fifth Workshop on Grand Unification}} (1984) }.

\bibitem{Slansky:1981yr}
R.~Slansky, \emph{{Group Theory for Unified Model Building}}, {\emph{Phys.
  Rept.} {\bfseries 79} (1981) 1}.

\bibitem{Feger_2020}
R.~Feger., T.~Kephart and R.~Saskowski, \emph{{LieART 2.0 -- A Mathematica
  application for Lie Algebras and Representation Theory}},
  \href{https://doi.org/10.1016/j.cpc.2020.107490}{\emph{Comput. Phys. Commun.}
  {\bfseries 257} (2020) 107490},
  [\href{https://arxiv.org/abs/1912.10969}{{\ttfamily 1912.10969}}].

\bibitem{D_Auria_2001}
R.~D'Auria, S.~Ferrara, M.~A. Lledo and V.~S. Varadarajan, \emph{{Spinor
  algebras}}, \href{https://doi.org/10.1016/S0393-0440(01)00023-7}{\emph{J.
  Geom. Phys.} {\bfseries 40} (2001) 101},
  [\href{https://arxiv.org/abs/hep-th/0010124}{{\ttfamily hep-th/0010124}}].

\bibitem{majoranaspinors}
J.~Figueroa-O'Farrill, ``{Lecture notes on Majorana Spinors}.'' School of
  Mathematics, University of Edinburgh, 2004.

\bibitem{Chatzistavrakidis:2010xi}
A.~Chatzistavrakidis, H.~Steinacker and G.~Zoupanos, \emph{{Orbifolds, fuzzy
  spheres and chiral fermions}},
  \href{https://doi.org/10.1007/JHEP05(2010)100}{\emph{JHEP} {\bfseries 05}
  (2010) 100}, [\href{https://arxiv.org/abs/1002.2606}{{\ttfamily 1002.2606}}].

\bibitem{Chatzistavrakidis:2011toc}
A.~Chatzistavrakidis, H.~Steinacker and G.~Zoupanos, \emph{{Orbifold matrix
  models and fuzzy extra dimensions}},
  \href{https://doi.org/10.22323/1.155.0047}{\emph{PoS} {\bfseries CORFU2011}
  (2011) 047}, [\href{https://arxiv.org/abs/1204.6498}{{\ttfamily 1204.6498}}].

\bibitem{ParticleDataGroup:2022pth}
{\scshape Particle Data Group} collaboration, R.~L. Workman et~al.,
  \emph{{Review of Particle Physics}},
  \href{https://doi.org/10.1093/ptep/ptac097}{\emph{PTEP} {\bfseries 2022}
  (2022) 083C01}.

\bibitem{King:2021gnh}
S.~F. King, S.~Pascoli, J.~Turner and Y.-L. Zhou, \emph{{Confronting SO(10)
  GUTs with proton decay and gravitational waves}},
  \href{https://doi.org/10.1007/JHEP10(2021)225}{\emph{JHEP} {\bfseries 10}
  (2021) 225}, [\href{https://arxiv.org/abs/2106.15634}{{\ttfamily
  2106.15634}}].

\bibitem{Coleman1967}
S.~Coleman and J.~Mandula, \emph{{All Possible Symmetries of the $S$ Matrix}},
  {\emph{Phys. Rev.} {\bfseries 159} (1967) 1251}.

\bibitem{StefasZoupanos:inprep}
S.~Stefas and G.~Zoupanos. in preparation, 2026.

\bibitem{NandiStefasZoupanos:inprep}
P.~Nandi, S.~Stefas and G.~Zoupanos. in preparation, 2026.

\bibitem{Stelle:1977ry}
K.~S. Stelle, \emph{{Classical Gravity with Higher Derivatives}},
  \href{https://doi.org/10.1007/BF00760427}{\emph{Gen. Rel. Grav.} {\bfseries
  9} (1978) 353}.

\bibitem{Starobinsky:1980te}
A.~A. Starobinsky, \emph{{A New Type of Isotropic Cosmological Models Without
  Singularity}},
  \href{https://doi.org/10.1016/0370-2693(80)90670-X}{\emph{Phys. Lett. B}
  {\bfseries 91} (1980) 99--102}.

\bibitem{hell2023degrees}
A.~Hell, D.~Lust and G.~Zoupanos, \emph{{On the degrees of freedom of R$^{2}$
  gravity in flat spacetime}},
  \href{https://doi.org/10.1007/JHEP02(2024)039}{\emph{JHEP} {\bfseries 02}
  (2024) 039}, [\href{https://arxiv.org/abs/2311.08216}{{\ttfamily
  2311.08216}}].

\bibitem{Boulware:1972yc}
D.~G. Boulware and S.~Deser, \emph{{Can gravitation have a finite range?}},
  \href{https://doi.org/10.1103/PhysRevD.6.3368}{\emph{Phys. Rev. D} {\bfseries
  6} (1972) 3368--3382}.

\bibitem{deRham:2010kj}
C.~de~Rham and G.~Gabadadze, \emph{{Generalization of the Fierz-Pauli Action}},
  \href{https://doi.org/10.1103/PhysRevD.82.044020}{\emph{Phys. Rev. D}
  {\bfseries 82} (2010) 044020},
  [\href{https://arxiv.org/abs/1007.0443}{{\ttfamily 1007.0443}}].

\bibitem{deRham:2010tw}
C.~de~Rham, G.~Gabadadze and A.~J. Tolley, \emph{{Resummation of Massive
  Gravity}}, \href{https://doi.org/10.1103/PhysRevLett.106.231101}{\emph{Phys.
  Rev. Lett.} {\bfseries 106} (2011) 231101},
  [\href{https://arxiv.org/abs/1011.1232}{{\ttfamily 1011.1232}}].

\bibitem{SchmidtMay:2015vnx}
A.~Schmidt-May and M.~von Strauss, \emph{{Recent developments in bimetric
  theory}}, \href{https://doi.org/10.1088/1751-8113/49/18/183001}{\emph{J.
  Phys. A} {\bfseries 49} (2016) 183001},
  [\href{https://arxiv.org/abs/1512.00021}{{\ttfamily 1512.00021}}].

\bibitem{Snyder:1946qz}
H.~S. Snyder, \emph{Quantized space-time},
  \href{https://doi.org/10.1103/PhysRev.71.38}{\emph{Phys. Rev.} {\bfseries 71}
  (1947) 38}.

\bibitem{yang1947}
C.~N. Yang, \emph{On quantized space-time},
  \href{https://doi.org/10.1103/PhysRev.72.874}{\emph{Phys. Rev.} {\bfseries
  72} (1947) 874}.

\bibitem{kastrup_1966}
H.~A. Kastrup, \emph{Position operators, gauge transformations, and the
  conformal group}, \href{https://doi.org/10.1103/PhysRev.143.1021}{\emph{Phys.
  Rev.} {\bfseries 143} (1966) 1021}.

\bibitem{Heckman_2015}
J.~J. Heckman and H.~Verlinde, \emph{Covariant non-commutative space--time},
  \href{https://doi.org/10.1016/j.nuclphysb.2015.02.018}{\emph{Nucl. Phys. B}
  {\bfseries 894} (2015) 58--74}.

\end{thebibliography}\endgroup

\end{document}